\documentclass{article}
\usepackage{amssymb}
\usepackage{amsmath}
\usepackage{amsfonts}
\usepackage{hyperref}
\usepackage[parfill]{parskip}
\usepackage{graphicx}
\let\UnmodifSec=\section
\renewcommand{\section}{\setcounter{equation}{0}\UnmodifSec}

\newtheorem{lemma}{Lemma}[section]

\newtheorem{remark}{Remark}[section]

\def\w{{\cal W}}
\def\ww{{\cal W}'}
\def\www{W}
\def\whw{w}

\def\bC{{\bf C}}
\def\bR{{\bf R}}
\def\bN{{\bf N}}
\def\bZ{{\bf Z}}
\def\Im{\mathop{\rm Im}\nolimits}
\def\Re{\mathop{\rm Re}\nolimits}

\def\bCp{{\bf C}_+}
\def\bCm{{\bf C}_-}
\def\tg{\mathop{\rm tg}\nolimits}
\def\th{\mathop{\rm th}\nolimits}
\def\ch{\mathop{\rm cosh}\nolimits}
\def\sh{\mathop{\rm sinh}\nolimits}

\def\CC{{\cal C}}
\def\DD{{\cal D}}

\def\FF{{\cal F}}

\def\HH{{\cal H}}

\def\LL{{\cal L}}

\def\NN{{\cal N}}

\def\SS{{\cal S}}
\def\TT{{\cal T}}
\def\UU{{\cal U}}

\def\XX{{\cal X}}

\def\wh{\widehat}
\def\wt{\widetilde}
\def\ovl{\overline}

\def \vhi{\varphi}
\def \veps{\varepsilon}

\def\half{{\scriptstyle{1 \over 2}}}
\def\interior#1{\setbox1=\hbox{$#1$}\rlap{$#1$}\kern0.4\wd1\raise1.1\ht1%
\hbox{$\scriptstyle \circ$}}
\def\bydef{\mathrel{\buildrel \hbox{\scriptsize \rm def} \over =}}
\def\boxit#1#2{\setbox1=\hbox{\kern#1{#2}\kern#1}%
\dimen1=\ht1 \advance \dimen1 by #1 \dimen2=\dp1 \advance \dimen2 by #1
\setbox1=\hbox{\vrule height\dimen1 depth\dimen2\box1\vrule}%
\setbox1=\vbox{\hrule\box1\hrule}%
\advance \dimen1 by .4pt \ht1=\dimen1 \advance \dimen2 by .4pt \dp1=\dimen2
\box1\relax}
\def\endprf{\raise .5ex\hbox{\boxit{2pt}{\ }}}

\def\Xcd{X_d^{(c)}}
\def\Xdc{X_d^{(c)}}

\def\ifundefined#1{\expandafter\ifx\csname#1\endcsname\relax}

\def\beq{\begin{equation}}
\def\endq{\end{equation}}
\def\beqa{\begin{eqnarray}}
\def\endqa{\end{eqnarray}}

\def\y{{\mathrm y}}

\def\k{\kappa}
\def\coupl{{\gamma}}
\def\kr{{\mathrm k}}

\begin{document}
\title{Particle decays and stability on the de Sitter universe}

\author{Jacques Bros$^1$, Henri Epstein$^2$ and Ugo Moschella$^3$ \\
$^1$Service de Physique th\'eorique - CEA. Saclay.
91191 Gif-sur Yvette.\\$^2$Institut des Hautes \'Etudes
Scientifiques, 91440 Bures-sur-Yvette.\\$^3$Universit\`a
dell'Insubria, Como and INFN Milano}
\maketitle

\begin{abstract}
We study particle decay in  de Sitter space-time as given by first
order perturbation theory in a Lagrangian interacting quantum field theory.
We study in detail the adiabatic limit of the perturbative amplitude
and compute the ``phase space'' coefficient exactly in the case
of two equal particles produced in the disintegration.
We show that for fields with masses
above a critical mass $m_c$ there is no such thing as particle
stability, so that decays forbidden in flat space-time do occur
here. The lifetime of such a particle also turns out to be
independent of its velocity when that lifetime is comparable with de
Sitter radius. Particles with mass lower than critical have a completely
different behavior:
the masses of their decay products must obey quantification rules,
and their lifetime is zero.
\end{abstract}
 \maketitle

\section{Introduction}

Some important progress in the astronomical observations of the last
ten years \cite{Riess:1998cb,Perlmutter:1998np} have led in a progressively convincing way
to the surprising conclusion that the recent universe is dominated
by an almost spatially homogeneous exotic form of energy density to
which there corresponds an effective negative pressure. Such
negative pressure acts repulsively at large scales, opposing itself
to the gravitational attraction. It has become customary to
characterize such energy density by the term "dark".

The simplest and best known candidate for the "dark energy" is the
cosmological constant. As of today, the $\Lambda$CDM (Cold Dark
Matter) model, which is obtained by adding a cosmological constant
to the standard model, is the one which is in better agreement with
the cosmological observations, the latter being progressively more
precise. Recent data show that dark energy behaves as a cosmological
constant within a few percent error. In addition, if the description
provided by the $\Lambda$CDM model is correct, Friedmann's equation
shows that the remaining energy components must in the future
progressively thin out and eventually vanish thus letting the
cosmological constant term alone survive.

In the above scenario the de Sitter geometry \cite{desitter,desitterbis}, which is the
homogeneous and isotropic solution of the vacuum Einstein equations
with cosmological term, appears to take the double role of reference
geometry of the universe, namely the geometry of spacetime deprived
of its matter and radiation content and of  geometry that the
universe approaches asymptotically.
On the other hand,  it seems reasonable to imagine that the presence
of a small cosmological constant, while having a huge impact on our
understanding of the universe as a whole, would not influence
microphysics in its quantum aspects. However this conclusion may
have to be reassessed, because in the presence of a cosmological
constant, however small, it is the notion of elementary particle
itself which has to be reconsidered: indeed, the usual asymptotic
theory is based on concepts which refer closely to the global
structure of Minkowski spacetime and to its Fourier representation,
and do not apply to the de Sitter universe which is not
asymptotically flat. Secondly, even if one may think that
interactions between elementary particles happen in a "laboratory"
so that "infinity" is a distance of the order of meters, our present
understanding of perturbative quantum field theory is also based on
global concepts; in particular, the calculation of perturbative
amplitudes involves integrations over the whole spacetime manifold and
it should be expected that different topological global structures
result in different physical properties in the "small".

The literature about de Sitter quantum field theory is very extensive,
but there is no comparison with the understanding one has of Minkowskian
field theory as regards both general and structural results as a
well as its operative and computational possibilities. This second
point is particularly doleful: calculations of perturbative
amplitudes which in the Minkowskian case would be simple or even
trivial become rapidly prohibitive or impossible in the case of de
Sitter or anti de Sitter universes: this in spite of the fact that
one is dealing with maximally symmetric manifolds which have
invariance groups of the same dimension as the Poincar\'e group. The
technical, but also the physical, difference lies precisely in 
the above mentioned fact that
much of the usual quantum field theory is based on concepts which
are characteristic of the global structure of Minkowski spacetime and 
which do not persist
in the presence of curvature, already in the
presence of a mere cosmological constant, where Minkowskian
spacetime is replaced by the de Sitter or by the anti de Sitter one.

In this paper we give a full description of how to solve the problem
of  calculating the mean lifetime of unstable scalar particles on
de Sitter spacetime at first order in perturbation theory.
This  interesting physical
problem  provides also an example of a concrete perturbative
calculation in presence of the cosmological constant. The task
already presents considerable mathematical difficulties.

To our knowledge this calculation was first taken up by
O.~Nachtmann \cite{N} in 1968.
He showed, in a very special case,
that while a Minkowskian particle can never decay into
heavier products, a dS-particle can, although this effect is
exponentially small in the dS-radius.

The subject has  acquired a greater physical interest with the advent of 
inflationary cosmology.
In particular, the idea that particle decays during the (quasi-)de Sitter phase
may have important consequences on the physics of the early universe has been 
suggested recently \cite{Boyanovsky:1996ab,Boyanovsky:2004gq,Boyanovsky:2004ph}.
The mathematical and physical difficulties related to
the lack of time-translation symmetry of the de Sitter universe,
and more generally of non-static cosmological backgrounds,
have been tackled \cite{Boyanovsky:1996ab,Boyanovsky:2004gq,Boyanovsky:2004ph}
by using the Schwinger-Keldysh formalism,
which is suitable for studying certain aspects of the
quantum dynamics of systems out of equilibrium.
An important ingredient of this approach is the so called 
Dynamic Renormalization Group
\cite{Boyanovsky:2003ui} which allows a kind of resummation of an 
infinite series of infrared diverging quantities.
That method is however based on the introduction of a practical notion of
lifetime of an unstable particle which is quite different
from the definition commonly used in quantum physics.
Also, the hard technical difficulties
of the concrete calculation involved in solving a complicated 
integro-differential equation
have only been faced in the favorably special conformal and 
minimally coupled massless cases although in principle the method can be 
used to deal with particles of generic mass
  \cite{Boyanovsky:1996ab,Boyanovsky:2004gq,Boyanovsky:2004ph}.

In this paper we perform a computation which is similar to the one 
outlined by Nachtmann and follows
the conventional quantum field theoretical  perturbative approach for computing
probability amplitudes. Our work gives significantly
wider results w.r.t. \cite{N}, e.g. regarding the so-called adiabatic
limit, complementary-series-particles, and explicit expressions
of the relevant K\"all\'en-Lehmann weights. On the other side comparing our 
result with those of  
\cite{Boyanovsky:1996ab,Boyanovsky:2004gq,Boyanovsky:2004ph} 
is not easy because of the non standard  (but interesting)
definition of lifetime chosen in 
\cite{Boyanovsky:1996ab,Boyanovsky:2004gq,Boyanovsky:2004ph}.

These findings  have been summarized in a recent
short communication \cite{Bros:2006gs}.
The results exhibit significant differences compared to the
Minkowski case, and decay processes which are normally forbidden
become possible
and, vice-versa, processes that are normally possible are now
forbidden. The maximal symmetry of the de Sitter universe implies
the existence of a global square-mass operator, one of the two Casimir
operators of the de Sitter group $SO_0(1, d)$ (see e.g.
\cite{Gursey}); this quantity is conserved for de Sitter invariant
field theories. However, in contrast with the Poincar\'e group case,
the tensor product of two unitary irreducible representations of
masses $m_1$ and $m_2$ decomposes into a direct integral of
representations whose masses $m$ do not satisfy the `subadditivity
condition' $m \ge m_1 + m_2$: all representations of mass larger than
a certain critical value (principal series) appear in the
decomposition. This fact was shown in \cite{N} for the
two-dimensional case and will be established here in general. This
means that the de Sitter symmetry does not prevent a particle with
mass in the principal series from decaying into e.g. pairs of
heavier particles. This phenomenon also implies that there can be
nothing like a mass gap in that range. This is a major obstruction
to attempts at constructing a de Sitter S-matrix; the Minkowskian
asymptotic theory makes essential use of an isolated point in the
spectrum of the mass operator, and this will generally not occur in
the de Sitter case. We will also show that the tensor product of two
representations of sufficiently small mass below the critical value
(complementary series) contains an additional finite sum of discrete
terms in the complementary series itself (at most one term in
dimension 4). This implies a form of particle stability, but the new
phenomenon is that a particle of this kind cannot disintegrate
unless the masses of the decay products have certain quantized
values. Stability for the same range of masses has also been
recently found [8] in a completely different context. Other remarks
about the physical meaning and applicability of our results will be
presented in the concluding section.

\subsection{Notation}

\label{nota} We denote $\bCp = - \bCm$ the open upper complex
half-plane. Let $\Delta = \bC \setminus [-1,\ 1]$, $\Delta_1 = \bC
\setminus (-\infty,\ 1]$. The function $\log$ is defined as
holomorphic on $\bC \setminus (-\infty,\ 0]$ and real on $(0,\
+\infty)$ and $\zeta \mapsto \zeta^\mu$ as $\exp(\mu\log (\zeta))$.
It is entire in $\mu$. If $\zeta \in \Delta_1$ and $\rho >0$, then
$(\rho\zeta)^\mu = \rho ^\mu\zeta^\mu$. If $\zeta \in \bCp$ and $s
\in \bCm$, then $(s\zeta)^\mu = s^\mu\zeta^\mu$. We define $z
\mapsto (z^2-1)^{1/2}$ as holomorphic on $\Delta$ and asymptotic to
$z$ at large $|z|$. It is Herglotz, negative on $(-\infty,\ -1)$ and
positive on $(1,\ +\infty)$.

\section{Free fields in Minkowski and de Sitter spacetimes}

In this section we give a short summary of the theory of free and
generalized free quantum fields on de Sitter spacetime.  Since there are infinitely many
inequivalent representations of the field algebra, a (mathematical) choice has to be made
on physical grounds.
Ours is based on the analyticity properties of the vacuum expectation
values: see the condition (W2) below.
In the Minkowski space,
this is equivalent to the positivity of the energy. In the de Sitter
case, it admits a thermal interpretation (\cite{Gibbons:1977mu,Bros:1994dn,Bros:1995js,Bros:1998ik}).
The reader can find in \cite{Bros:1994dn,Bros:1995js,Bros:1998ik} a general approach to
to de Sitter QFT  based on such analytic properties. It includes the so called Bunch-Davies,
also called Euclidean vacuum of de
Sitter scalar Klein-Gordon fields as a basic example.

The real (resp. complex) $d$-dimensional Minkowski spacetime $M_d$
(resp. $M_d^{(c)}$) is $\bR^d$ (resp. $\bC^d$) equipped
with the Lorentzian inner product
\begin{equation}
x\cdot {x'} = x^0 {x'}^0 -x^1 {x'}^1 - \ldots -x^{d-1}
{x'}^{d-1} = x^0 {x'}^0 - \vec{x}\cdot \vec{{x}}'\
\label{f.1}\end{equation}
w.r.t. an arbitrarily chosen Lorentz frame
$\{e_\sigma, \sigma = 0,\ldots, d-1\}$ .  When no ambiguity arises,
$x^2 \bydef x\cdot x$. The real (resp. complex) de Sitter spacetime
$X_d$ (resp. $X_d^{(c)}$) with radius $R > 0$ are the hyperboloids
\begin{equation}
X_d = \{x \in M_{d+1}\ :\ x\cdot x + R^2 = 0\},\ \ \ \
X_d^{(c)} = \{x \in M_{d+1}^{(c)}\ :\ x\cdot x + R^2 = 0\},
\label{f.2}\end{equation}
equipped with the pseudo-riemannian metric
induced by (\ref{f.1}). $L_+^{\uparrow}(d) = SO(1,\ d-1;\ \bR)$ is
the connected Lorentz group acting on $M_d$, and $L_+(\bC;\ d) =
SO(1,\ d-1;\ \bC)$ is the connected complex Lorentz group acting on
$M_d^{(c)}$. The connected group of displacements on $X_d$ (resp.
$X_d^{(c)}$) is $L_+^{\uparrow}(d+1)$ (resp. $L_+(\bC;\ d+1)$),
sometimes denoted $G_0$ (resp. $G_0^{(c)}$). These groups act
transitively. Note that our definition of $M_d$ etc. arbitrarily
selects a particular orthonormal basis $(e_0,\ \ldots,\ e_{d-1})$ in
$M_d$ or $(e_0,\ \ldots,\ e_{d})$ in $M_{d+1}$. These particular
Lorentz frames will be useful in the sequel. In $M_d$ the future and
past open cones $V_\pm$ and the future and past light-cones $C_\pm$
are given by
\begin{eqnarray}
V_+ &=& \{x \in M_d\ :\ x\cdot x > 0,\
\ x^0 > 0\} = - V_-\ ,
\nonumber\\
C_+ &=& \{x \in M_d\ :\ x\cdot x = 0,\ \ x^0 \ge 0\} = - C_-\ .
\label{f.3}\end{eqnarray}
The future and past tubes in the complex
Minkowski spacetime $M_d^{(c)}$ are given by:
\begin{equation}
{\rm
T}_\pm = \bR^d +iV_\pm\ .
\label{f.4}\end{equation}
The future and
past tuboids in $X_d^{(c)}$ are the intersections of the future and
past tubes in $M_{d+1}^{(c)}$ with the complex de Sitter manifold
$X_d^{(c)}$:
\begin{equation}
\TT_\pm = {\rm T}_\pm \cap X_d^{(c)}\ .
\label{f.5}\end{equation} We
will use the letter $\XX$ to denote either $M_d$ or $X_d$ when the
same discussion applies to both, $\XX^{(c)}$ denoting the
complexified object. $dx$ will denote the standard invariant measure
on $\XX$, i.e. using the frame $(e_0,\ \ldots,\ e_n)$, $dx =
dx^0\ldots dx^{d-1}$ in the case of $M_d$, and $dx =
2\delta(x^2+R^2)\,dx^0\ldots dx^d$ for $X_d$.

A (neutral scalar) generalized free field $\phi$ on $\XX$  is
entirely specified by its 2-point function.
This is a tempered distribution
$\w$ on $\XX\times \XX$ (we denote $\ww(x,\ {x'}) = \w({x'},\ x)$),
which we require to have the following properties:
\begin{description}

\item{(W1) Hermiticity:}
\begin{equation}
\ovl{\w(x,\ {x'})} = \w({x'},\ x)\ .
\label{f.6}\end{equation}

\item{(W2) Analyticity and invariance:}
there is a function $\whw$ of one complex variable, holomorphic in
the cut plane $\bC \setminus \bR_+$, with tempered behavior at
infinity and at the boundaries, such that, in the sense of tempered
distributions,
\begin{equation} \w(x,\ {x'}) =
\lim_{\begin{array}{c}
z \in {\rm T}_-,\ z' \in {\rm T}_+\\
z \rightarrow x,\ z' \rightarrow {x'}
\end{array}} \whw ((z - {z'})^2),
\label{f.7}\end{equation}
hence
\begin{equation}
\ww(x,\ {x'}) =
\w({x'},x)= \lim_{\begin{array}{c}
z \in {\rm T}_+,\ z' \in {\rm T}_-\\
z \rightarrow x,\ z' \rightarrow {x'}
\end{array}} \whw ((z - z')^2)\ .
\label{f.7.1}\end{equation}
For complex $z,\ z' \in \Xdc$ such that
$(z-z')^2 \in \bC \setminus \bR_+$ we will denote $\www(z,\ z') =
\whw ((z - z')^2)$. Note that this implies
\begin{equation}
\www(z,\ z') = \www(z',\ z) = \www(-z',\ -z),
\label{f.7.2}\end{equation}
and
\begin{equation}
\w(x,\ {x'})  = \w(-{x'},\ -x)\ .
\label{f.7.3}\end{equation}
(W1) and (W2) also imply
\begin{equation} \www(z,\ z') = \ovl{\www(\ovl{z},\ \ovl{z}')}.
\label{f.7.4}\end{equation}

\item{(W3) Positivity:}
For every $f \in \SS(\XX)$,
\begin{equation}
\langle \w,\ \ovl{f}\otimes f \rangle =
\int_{\XX\times \XX} \ovl{f(x)}\,\w(x,\ {x'})\,f({x'})\,dx\,d{x'} \ge 0\ .
\label{f.8}\end{equation}
\end{description}

Conversely, given $\w$ and $w$ having these properties, after
having identified the kernel $\NN_1=\{f\in\SS(\XX)\ :\ \langle \w,\
\ovl{f}\otimes f \rangle = 0\}$  we can construct a Hilbert space
$\FF_1$ by completing $\SS(\XX)/\NN_1$ equipped with the scalar
product $(f,\ g) = \langle \w,\ \ovl{f}\otimes g \rangle$, and then
exponentiate $\FF_1$ into a Fock space $\FF$
\begin{equation} \FF =
\bigoplus_{n=0}^\infty \FF_n,\ \ \ \FF_0 = \bC,\ \ \ \FF_n =
S\FF_1^{\otimes n}\ \ {\rm for}\ \ n \ge 1\ .
\label{f.9}\end{equation}
The vacuum $\Omega$ is the unit vector $1
\in \FF_0 = \bC$. There is a continuous unitary representation $U$
of the Poincar\'e or de Sitter group acting on $\FF$ and preserving
the $\FF_n$, with $U\Omega = \Omega$. The generalized free field
$\phi$ is defined on a dense domain in $\FF$ and $(\Omega,\
\phi(x)\phi({x'})\Omega) = \w(x,\ {x'})$.

As a result of the analyticity property (W2), the Wick powers of a
generalized free field are well-defined local fields operating in
the same Fock space. Their vacuum expectation values are obtained by
the standard Wick formulae as sums of products of $\w$.

We note that a function $\w$ on $X_d\times X_d$ possessing the
properties (W1) and (W2) automatically extends (through (\ref{f.7}))
to a function with the same properties on $M_{d+1}\times M_{d+1}$,
so that a generalized free field on $X_d$ has an extension as a
generalized free field on $M_{d+1}$. However the extension of $\w$
need not satisfy (W3) on $M_{d+1}\times M_{d+1}$ even if it does on
$X_d\times X_d$.

A free field $\phi$ of mass $m >0$ on $\XX$ is a generalized free
field such that $\w$ is a solution of the Klein-Gordon equation with
mass $m$ in both arguments, and is normalized so as to obey the
canonical commutation relations. In that case $\w$ is uniquely
determined by $m$ and will be denoted $\w_{m}$. In the Minkowskian
case, the representation $U|\FF_1$ is irreducible and equivalent to
the representation $[m,\ 0]$ of the Poincar\'e group.

As usual, the representation $U$ provides a representation
of the Lie algebra of the (Poincar\'e or de Sitter) group
and its envelopping algebra
by self-adjoint (or $i\times$ self-adjoint) operators on
$\FF$. In particular the square-mass operator $M^2$ is
given by $M^2 = P^\mu P_\mu$ in the Minkowskian case,
and by $M^2 = M^{\mu\nu}M_{\mu\nu}/2R^2$ in the de Sitter case.
In both cases, $M^2 \Psi = m^2\Psi$ for every $\Psi\in\FF_1$.
See e.g. \cite{Gursey}.

\subsection{Special features of free fields in de Sitter space-time}

In the de Sitter case the mass $m$ can
be related to a dimensionless parameter $\nu$ as follows
\begin{equation}
m^2R^2 = \left ( {d-1 \over 2} \right )^2 + \nu^2\ ,
\label{f.10}\end{equation}
\begin{equation}
\nu = \pm \left [
m^2R^2 - \left ( {d-1 \over 2} \right )^2 \right ]^{1/2} =
\pm R(m^2-m_c^2)^{1/2},\ \ \ \ m_c = {d-1\over 2R}\ .
\label{f.11}
\end{equation}
In this case, if no ambiguity arises, we shall often denote
$\w_{\nu} = \w_{-\nu}$ to mean $\w_{m}$, and similarly $\www_\nu$
and $\whw_\nu$. Explicitly, if $z,\ z' \in \Xcd$, $(z-z')^2 \notin
\bR_+$,  and hence $\zeta = z\cdot z'/ R^2$ does not belong to the
real interval $-\infty, -1]$),
\begin{eqnarray} \www_\nu(z,z') &=& {\Gamma\left ({d-1\over
2}+i\nu \right ) \Gamma\left ({d-1\over 2}-i\nu \right )\over
2(2\pi)^{d\over 2}R^{d-2}}\, (\zeta^2-1)^{-{d-2\over 4}}\,
P^{-{d-2\over 2}}_{-\half+i\nu}(\zeta)
\label{f.12}\\
&=& {\Gamma\left ({d-1\over 2}+i\nu \right ) \Gamma\left ({d-1\over
2}-i\nu \right )\over (4\pi)^{d\over 2}R^{d-2}\Gamma\left({d\over 2}
\right )}\,F \left ( {d-1\over 2}+i\nu,\ {d-1\over 2}-i\nu\ ;\
{d\over 2}\ ;\ {1-\zeta \over 2} \right )\ . \label{f.13}
\end{eqnarray}
Since $\Gamma(c)^{-1}F(a,\ b\ ;\ c\ ;\ z)$ is entire in $a$, $b$,
and $c$ the rhs of (\ref{f.13}) is meromorphic in $\nu$ with simple
poles at $\nu = \pm i( (d-1)/2 + n)$, $n \ge 0$ an integer. In other
words $\w_\nu(z,z')$ extends to a holomorphic function of $\nu$, $z$
and $z'$ in the domain $\{\nu \in \bC,\ z\in \Xcd,\ z'\in \Xcd\ :\
\nu \notin \pm i((d-1)/2 + \bZ_+),\ \ (z-z')^2 \notin \bR_+\}$.
However $w_\nu$ possesses the positivity property (W3) (see
(\ref{f.8})) only if either
\begin{description}

\item{(1)}
$\nu$ is real, i.e. $m \ge m_c = (d-1)/2R$. In this case $U | \FF_1$ is
an irreducible unitary representation of the ``principal series''.

or

\item{(2)}
$\nu$ is pure imaginary with $i\nu \in (-(d-1)/2,\ (d-1)/2)$, i.e.
$0 < m \le (d-1)/2R$. In this case $U | \FF_1$ is an irreducible
unitary representation of the ``complementary series''.
\end{description}

We shall need a small part of the harmonic analysis on
the de Sitter space-time as developed in \cite{Bros:1995js}.
If $z \in \TT_\pm \subset X_d^{(c)}$ and
$\xi \in C_+\setminus \{0\} \subset M_{d+1}$, then $\pm \Im (z\cdot\xi) >0$,
so that $(z\cdot\xi)^\lambda$ is well-defined and holomorphic in $(z,\ \lambda)$
in $(\TT_+ \cup \TT_-)\times \bC$.
The role of plane waves on $X_d$ is played by the distributions
\begin{equation}
\psi_{\lambda}^{\pm} (x ,\xi) = \lim_{y \in V_{+},\ y \rightarrow 0}
((x\pm iy)\cdot \xi)^{\lambda} = \ovl{\psi_{\bar\lambda}^{\mp} (x
,\xi)}. \label{f.13.1}
\end{equation}
An important formula expressing the de Sitter case two-point
$\www_\nu$ as a Fourier superposition of plane-waves  is the
following (see \cite{Bros:1995js}):
\begin{equation}
\www_\nu(z,z') = R\, c_{d,\nu}\int_{\gamma}\left({z\cdot
\xi}\right)^{-{d-1\over 2} + i\nu} \left({\xi\cdot
z'}\right)^{-{d-1\over 2} - i\nu}\alpha(\xi) ,
\label{f.14}\end{equation} where $z_1 \in \TT_-$, $z_2 \in \TT_+$,
and
\begin{equation}
c_{d,\nu}=\,\,{\Gamma({d-1\over 2}+i\nu)\Gamma({d-1\over 2}-i\nu)
e^{-\pi\nu} \over 2^{ d+1} \pi^d}\, . \label{f.15}\end{equation} In
(\ref{f.14}), $\gamma$ is a ($d-1$)-cycle in $C_+\setminus \{0\}$
homologous to the sphere $S_0 = C_+ \cap \{\xi\ :\ \xi^0 = 1\}$. The
($d-1$)-form $\alpha$ is given, in the standard coordinates, by
\begin{equation}
\alpha = (\xi^0)^{-1} \sum_{j=1}^d (-1)^{j+1} \xi^j\,d\xi^1\ldots\
\wh{d\xi^j}\ldots\ d\xi^d\ . \label{f.16}\end{equation} If a smooth
function $f$ on $C_+\setminus \{0\}$ is homogeneous of degree
$(1-d)$, the form $f\alpha$ is closed, so that the linear functional
\begin{equation}
f \mapsto I_0(f) = \int_\gamma f(\xi)\alpha(\xi)
\label{f.16.1}\end{equation} is independent of $\gamma$. This
implies that it is Lorentz-invariant. We often denote $d\mu_\gamma$
the measure defined on $\gamma$ by the restriction of $\alpha$. In
particular the restriction of $\alpha$ to the $(d-1)$-sphere $S_0$
is the standard volume form on that sphere, normalized by
$\int_{S_0} d\mu_{S_0}(\xi) = {2 \pi^{d/2}/\Gamma(d/2)}$. It is
possible to take the limit of (\ref{f.14}), in the sense of
distributions, when $z_1$ and $z_2$ tend to the reals:
\begin{equation}
\w_{\nu}(x,\ x') = R\, c_{d,\nu}\int_{\gamma}
\psi_{-{d-1\over 2}+i\nu}^-(x,\xi)\,
\psi_{-{d-1\over 2}-i\nu}^+(x',\xi)\,d\mu_\gamma(\xi)\ .
\label{f.16.2}\end{equation}
Comparing (\ref{f.13}) with (\ref{f.14}) and (\ref{f.15}) gives
\begin{equation}
\int_{\gamma}\left({z\cdot \xi}\right)^{-{d-1\over 2} + i\nu}
\left({\xi\cdot z'}\right)^{-{d-1\over 2} - i\nu}d\mu_\gamma(\xi) =
{e^{\pi\nu} 2\pi^{d/2}\over R^{d-1} \Gamma \left ({d\over 2}\right
)} F \left ( {d-1\over 2}+i\nu,\ {d-1\over 2}-i\nu\ ;\ {d\over 2}\
;\ {1-\zeta \over 2} \right )\ . \label{f.17}\end{equation} Both
sides of this equation are holomorphic in $z_1,\ z_2,\ \nu$ in the
domain $\TT_-\times\TT_+\times\bC$, hence the equation (\ref{f.17})
holds in this domain.

\begin{remark}\label{homg}\rm
If $T$ is a homogeneous distribution of degree $\beta$
on $C_+\setminus\{0\}$, it can be restricted to any $\CC^\infty$
submanifold of dimension $d-1$ which is transversal to the generators of
$C_+$, in particular to hyperplanar sections such as
$S_0= \{\xi \in C_+\ :\ \xi^0=1\}$ and
$V_0= \{\xi \in C_+\ :\ \xi^0+\xi^d = 1\}$. If $\gamma$ is of this type
and compact, $\int_\gamma T(\xi)\alpha(\xi)$ is well-defined and,
if $\beta=1-d$, it is independent of $\gamma$.
\end{remark}

\begin{remark}\label{psipm}\rm
For any complex $\alpha$, $(z,\ \xi) \mapsto (z\cdot \xi)^\alpha$
is $\CC^\infty$ in $\xi$ and holomorphic in $z$ on
$\TT_\pm \times (C_+\setminus \{0\})$ and it is an entire
function of $\alpha$. For each $\xi$ it has a limit in the sense of
tempered distributions on $X_d$ as $z$ tends to the reals, and this
has been denoted $\psi_\alpha^\pm(x,\ \xi)$. It is an entire
function in $\alpha$. Furthermore its invariance under $G_0$ implies
that, if $\vhi \in \SS( C_+\setminus \{0\})$,
$\int_{C_+} \psi_\alpha^\pm(x,\ \xi)\vhi(\xi)\,d\xi$ is $\CC^\infty$ in $x$.
Indeed any small displacement of $x$ can be effected by a group
transformation close to the identity, which can be transferred to
$\xi$ and thence to $\vhi$. In the same way, $\psi_\alpha^\pm(x,\ \xi)$
is $\CC^\infty$ in $\xi$ (as well as homogeneous)
when integrated with a smooth test-function in $x$.
This explains the meaning of formulae such as (\ref{f.16.2}).
Note that the integral in this formula is entire in $\nu$.
For similar reasons, for any $\vhi \in \SS(X_d)$,
$\int_{X_d}\vhi (x)\,\w_\nu(x,\ x')\,dx$
is $\CC^\infty$ in $x'$ and meromorphic in $\nu$.

\end{remark}

\subsection{More features common to Minkowski and de Sitter space-time}

An important formula, which holds in Minkowski as well as in
de Sitter space-time (but in this case only if $m,\ m' \ge m_c$),
is the {\it projector identity}:
\begin{equation}
\int_\XX \w_{m}(z,\ x)\,\w_{m'}(x,\ y)\,dx =
C_1(m,\ d)\delta(m^2-m^{\prime 2}) \w_{m}(z,\ y).
\label{f.18}\end{equation}
Here
\begin{eqnarray}
C_1(m,\ d) &=& 2\pi\ \hbox{for Minkowski space-time},\\
C_1(m,\ d) &=& C_0(\nu) = 2\pi |\coth(\pi\nu)|\ \hbox{for de Sitter
space-time}. \label{f.19}\end{eqnarray} The proof of the above
identity is trivial in the Minkowskian case. For the de Sitter case it
will be provided in Appendix \ref{projector}. Note that $C_0(mR)$
tends to $2\pi$ as $R \rightarrow +\infty$ for a fixed $m>0$.

The {\it K\"all\'en-Lehmann decomposition theorem} exists in both
$M_d$ and $X_d$. In the case of $M_d$, (see \cite{BLOT}, p.~360), it
asserts that, for every $\www$ having the properties (W1) and (W2)
there is a tempered $\rho$ such that
\begin{equation}
\www(z,\ z') = \int_{\bR_+} \rho(m^2)\,\www_m(z,\ z')\,dm^2\ .
\label{f.20}\end{equation} If $\w$ satisfies (W3), then $\rho$ is a
tempered positive measure. The same holds in the dS case provided $
\www$ satisfies some decrease property. In this case, the integral
runs on masses of the principal series, i.e. $m > m_c = (d-1)/2R$.
For proofs and details, see \cite{Bros:1995js,bv}. In particular if $m_j
\ge 0$ and, in the dS case, $m_j > m_c$ for $1\le j \le N$,
\begin{equation}
\prod_{j=1}^N \w_{m_j}(x,\ x') = \int_{a \ge b}^\infty \rho(a^2;\
m_1,\ldots,\ m_N)\,\w_{a}(x,\ x')\,da^2. \label{f.21}\end{equation}
Here $b=m_c$ in the de Sitter case, $b = \sum_j m_j$ in the
Minkowski case.

\section{Particle decays: general formalism}

There is at the moment nothing like the Haag-Ruelle asymptotic
theory (HRT) (see \cite{jost, araki, BLOT}) for the de Sitter universe.
Indeed all the ingredients of that theory are missing in the de
Sitter case. For example, as it will be shown in this paper, even in
a free field theory of mass $m
> m_c$, the mass $m$ is not an isolated point in the mass spectrum.
Moreover the solutions of the Klein-Gordon equation do not have the
kind of localization at infinity which plays an essential role in
the HRT. The concept of a particle is therefore not obvious in de
Sitter space-time, except for localized observations. Here we adopt
Wigner's point of view: a one-particle vector state is a state
belonging to an invariant subspace of the Hilbert space in which the
representation of the invariance group reduces to an irreducible
representation. In the dS case, we also require that this
irreducible representation belong to the principal or complementary
series, i.e. it should be equivalent to one of the representations
which occur in the $\FF_1$ of a free field.

We shall study the decay of a particle using first-order
perturbation theory. The initial framework and calculations are the
same for the Minkowski and de Sitter cases: its ingredients are the
projector identity and the K\"all\'en-Lehmann representation.
(It can also be extended to the Minkowskian thermal case (\cite{Bros:1996mw})
although there is no K\"all\'en-Lehmann representation there).
Let
\begin{equation}
\phi_0,\ \phi_1,\ \ldots,\ \phi_N
\label{p.1}\end{equation}
be $1+N$ independent free scalar fields with masses $m_0 > 0,\ m_1>0,
\ldots,\ m_N>0$, acting in a common Fock space $\HH$, the tensor
product of the individual Fock spaces for the $\phi_k$\ :
\begin{equation}
\HH = \bigotimes_{k=0}^N \FF^{(k)}\ , \label{p.1.1}\end{equation}
\begin{equation}
(\Omega,\ \phi_j(x)\,\phi_k(y)\Omega) =
\delta_{jk}\,\w_{m_j}(x,\ y)\ .
\label{p.2}\end{equation}
We denote
\begin{equation}
\HH_{j_0,\ldots,j_N} = \FF_{j_0}^{(0)}\otimes \ldots \otimes
\FF_{j_N}^{(N)}\ . \label{p.2.1}\end{equation} This is the subspace
of states in $\HH$ containing $j_k$ $k$-particles.
$E_{j_0,\ldots,j_N}$ denotes the hermitian projector onto this
subspace. We now switch on an interaction term
\begin{equation}
\int_\XX \coupl\,g(x)\,{\LL}(x)\,dx,\ \ \
{\LL}(x) =\ :\phi_0(x)\phi_1(x)^{q_1}\ldots\phi_N(x)^{q_N}:\ .
\label{p.3}\end{equation}
Here the $q_j$ are non-negative integers,
and we denote $q! = \prod_{j=1}^N q_j!$. $\coupl$ is a small
constant. $g$ is a smooth, rapidly decreasing function over $\XX$.
In the end, $g$ should be made to tend to 1 (adiabatic limit).
According to perturbation theory, the transition amplitude between
two normalized states $\psi_0$ and $\psi_1$ in $\HH$ is given by
$(\psi_0,\ S(\coupl g) \psi_1)$, where $S(\coupl g)$ is the formal
series in $\coupl g$
\begin{equation}
S(\coupl g) = \sum_{n=0}^\infty {i^n\coupl^n \over n!} \int_{\XX^n}
g(x_1)\,dx_1\cdots g(x_n)\,dx_n\, T(\LL(x_1)\ldots \LL(x_n))\ \
\label{p.5}\end{equation}
In (\ref{p.5}), $T(\LL(x_1)\ldots \LL(x_n))$
denotes the (renormalized) time-ordered product of
$\LL(x_1),\ldots \LL(x_n)$. In the first order in $\coupl g$, the
transition amplitude between two orthogonal states $\psi_0$ and
$\psi_1$ is
\begin{equation}
(\psi_0,\ iT_1(\coupl g)\psi_1),\ \ \ \ T_1(\coupl g) = \int_\XX
\coupl g(x)\,\LL(x)\,dx\ .
\label{p.7}\end{equation}
We take
\begin{eqnarray}
\psi_0 &=& \int f_0(x)\,\phi_0(x)\Omega\,dx\ ,
\label{p.8}\\
\psi_1 &=& \int f_1(x_{11},\ldots,x_{1q_1},\ \ldots,\
x_{N1},\ldots,x_{Nq_N}) :\prod_{j=1}^N \prod_{k=1}^{q_j}
\phi_j(x_{jk})\,dx_{jk}:\,\Omega\ ,
\label{p.9}\end{eqnarray}
where
$f_0$ and $f_1$ are smooth rapidly decreasing functions. The states
of the form (\ref{p.8}) generate $\HH_{1,0\ldots,0}$ and the states
of the form (\ref{p.9}) generate $\HH_{0,q_1,\ldots,q_N}$. The
probability of transition from $\psi_0$ to any state in
$\HH_{0,q_1,\ldots,q_N}$ is:
\begin{eqnarray}
\Gamma =  { (\psi_0,\ T_1(\coupl g)E_{0,q_1,\ldots,q_N} T_1(\coupl
g)^*\,\psi_0)\over (\psi_0,\ \psi_0)} =  {q!\coupl^2\over (\psi_0,\
\psi_0)}\int \ovl{f_0(x)}\,f_0(y)\,g(u)\,g(v) \, \times \cr \times\
\w_{m_0}(x,\ u)\, \left \{ \prod_{j=1}^N \w_{m_j}(u,\ v)^{q_j}
\right \} \w_{m_0}(v,\ y)\, dx\,du\,dv\,dy\ .
\label{p.11}\end{eqnarray}
{}From now on, we suppose, in the dS case,
that $m_k > m_c$, $0 \le k \le N$, i.e all particles belong to the
principal series. We may then replace the central two-point function
in $u$ and $v$ by its K\"all\'en-Lehmann decomposition:
\begin{equation}
\prod_{j=1}^N \w_{m_j}(u,\ v)^{q_j} = \int
\rho(a^2;m_1,\ldots,m_1,\ldots,m_N,\ldots,m_N)\,\w_{a}(u,\ v)\,da^2\ .
\label{p.12}\end{equation}
Here $m_j$ occurs $q_j$ times as an
argument of $\rho$. This gives
\begin{eqnarray}
\Gamma = \frac{q!\coupl^2}{(\Psi_0,\Psi_0)}\,\int
\ovl{f_0(x)}\,f_0(y)\,g(u)\,g(v)\,
\rho(a^2;m_1,\ldots,m_1,\ldots,m_N,\ldots,m_N) \ \times \cr \times \
\w_{m_0}(x,\ u)\,\w_a(u,\ v)\,\w_{m_0}(v,\ y)\,
dx\,du\,dv\,dy\,da^2. \label{p.13}
\end{eqnarray}
The next step would  be the so-called adiabatic limit, and should
consist in letting the cut-off $g$ tend to 1 in this formula. It is
however easier to set first only one of the $g$'s equal to 1, say
$g(u)=1$ in (\ref{p.13}). It then becomes possible to perform the
integration over $u$ by using the projector identity (\ref{f.18})
and  we find for the transition probability:
\begin{equation}
\Gamma=L_1(f_0,\ g) \times\,
q!\,\rho(m_0^2;m_1,\ldots,m_1,\ldots,m_N,\ldots,m_N)\ ,
\label{decayy}\end{equation}
where
\begin{equation}
L_1(f_0,\ g) =
{\coupl^2\,C_1(m_0,\ d)\, \int g(v)\ovl{f_0(x)}\w_{m_0}(x,\
v)\w_{m_0}(v,\ y)f_0(y)dx\,dy\,dv\over \int \ovl{f_0(x)}\w_{m_0}(x,\
y)\,f_0(y)\,dx\,dy}\ .
\label{p.16}\end{equation}
This formula exhibits an interesting factorization: the first
factor depends only on the wavepacket $f_0$, the mass $m_0$ of the
incoming particle and the switching-off factor $\coupl^2g$; the
adiabatic limit still remains to be done there;  the second factor
contains all the information about the decay products.

If we now attempt to set $g(v) =1$ in (\ref{p.16}) and to integrate
over $v$ using again (\ref{f.18}), the result is proportional to
$\delta(m_0^2-m_0^2)$, i.e. the integral diverges. This difficulty
was resolved in the 1930's by aiming at the average transition
probability per unit time (see e.g. \cite{Veltman}, pp.~60-62). We first
review the well-known  Minkowski case, in a form which can serve as
a model for the de Sitter case. In fact even this famous old case
deserves some re-examination on its own right and it is possible,
in this case, to allow the
two $g$ in (\ref{p.13}) to tend to 1 simultaneously, or even at
different rates. This is done in Appendix \ref{admink}. It is found that, if
both $g$ are taken as in (\ref{p.20}), the result is the same as
found above. But this is not necessarily the case for other $g$.
Nevertheless the procedure announced above (i.e. setting
the first $g$ in (\ref{p.13}) be equal to 1, then discussing
the time average of the limit as the second $g$ tends to 1)
will be used in the de Sitter case, since it gives good results
in the Minkowski case, and
since calculations in the dS case would become much more difficult
otherwise.
Note that in the de Sitter case (\ref{decayy}) and (\ref{p.16})
are applicable only when $m_0 > m_c$ and the range of integration
over $a^2$ in (\ref{p.13}) contains only values $a^2 > m_c^2$
($m_c =(d-1)/2/R$). In the case of the decay into two particles of mass $m_1$,
it will be seen below that this includes the case $m_1 > m_c$, but also
the case $m_c > m_1 > m_c\sqrt{3}/2$.

\section{Minkowski case}
\subsection{Adiabatic limit: the Fermi golden rule}
The simplicity of the Minkowskian case arises from being able
to  use of the Fourier representations:
\begin{equation}
f_0(x) = \int e^{-ipx}\wt f_0(p)dp,\ \ \ g(x) = \int e^{-ipx}\wt
g(p)dp,\ \ \ w_m(x,\ y) = (2\pi)^{1-d}\int e^{ip(y-x)}\,\delta(p^2 -
m^2)\theta(p^0)\,dp. \label{p.17}\end{equation} Then the factor in
(\ref{p.16}) becomes
\begin{equation}
L_1(f_0,\ g) = {(2\pi)^{2}\coupl^2 \int \ovl{\wt
f_0(p)}\,\delta(p^2-m_0^2)\theta(p^0) \wt
f_0(q)\,\delta(q^2-m_0^2)\theta(q^0)\,\wt g(p-q)\, dp\,dq\over \int
|\wt f_0(p)|^2\,\delta(p^2-m_0^2)\theta(p^0)\,dp}\ .
\label{p.18}\end{equation} We now specialize the cut-off $g$ to
depend only on the time coordinate of the chosen frame $g(v) =
h(v^0)=h(t)$, i.e. we think of the interaction as smoothly switched
on and then turned off. The Fourier representation is then $ \wt
g(p) = \wt h(p^0)\,\delta(\vec{p})\  \label{p.19} $ and eq.
(\ref{p.18}) becomes
\begin{equation}
L_1(f_0,\ g) = {(2\pi)^{2}\coupl^2\, \wt h(0) \int (2p^0)^{-1}\,|\wt
f_0(p)|^2\,\,\delta(p^2-m_0^2)\theta(p^0)\,dp \over\int |\wt
f_0(p)|^2\,\delta(p^2-m_0^2)\theta(p^0)\,dp}\ .
\label{p.20}\end{equation}
If we choose for $g$ the indicator
function of a time-slice of thickness $T$,
i.e. $h(t) = \theta(t+T/2)\,\theta(T/2
-t)$, $\wt h(0) = T/2\pi$, we get
\begin{equation}
 L_1(f_0,\ g) = \ T\,\times \left({(2\pi)\coupl^2\, \int
(2p^0)^{-1}\,|\wt f_0(p)|^2\,\,\delta(p^2-m_0^2)\theta(p^0)\,dp
\over\int |\wt f_0(p)|^2\,\delta(p^2-m_0^2)\theta(p^0)\,dp}\right)\ .
\label{p.22}\end{equation}

Therefore, as noted above, removing the cut-off
produces infinity. However, according to the Fermi golden
rule, what is physically meaningful is not the amplitude but the
amplitude per unit time. Therefore, dividing this by $T$
and taking the limit as $T\rightarrow \infty$ (a particularly
trivial operation in this case) we finally get the following
expression for the transition probability per unit time:
\begin{equation}
\frac{1}{\tau(f_0)} = {(2\pi)\coupl^2\, \int
(2p^0)^{-1}\,|\wt f_0(p)|^2\,\,\delta(p^2-m_0^2)\theta(p^0)\,dp
\over\int |\wt f_0(p)|^2\,\delta(p^2-m_0^2)\theta(p^0)\,dp}
\,q!\,\rho(m_0^2;m_1,\ldots,m_1,\ldots,m_N,\ldots,m_N).
\label{p.23}\end{equation}
The reciprocal of this expression is the
lifetime of the 0-particle in the state $f_0$. { The dependence on
the wavepacket $f_0$ is a crucial feature of the special
relativistic Minkowski}{ case as it will be readily recognized}. For
instance to compute the lifetime $\tau_0$ of a particle at rest in
the chosen frame we may let $|\wt f_0(p)|^2$ tend to
$\delta(\vec{p})$, e.g. by taking
\begin{equation}
\wt f_0(p) = \veps^{(1-d)/2}\wt
\vhi(\vec{p}/\veps),\ \ \ f_0(x) = 2\pi\delta(x_0)
\veps^{(d-1)/2}\vhi(\veps \vec{x}),\ \ \ \veps > 0,
\label{p.24}\end{equation}
with $\wt \vhi \in \SS(\bR^{d-1})$, and
letting $\veps \rightarrow 0$. Then (\ref{p.23}) tends to
\begin{equation}
\frac{1}{\tau_0} = {\pi\coupl^2\over
m_0}\,q!\,\rho(m_0^2;m_1,\ldots,m_1,\ldots,m_N,\ldots,m_N)\ .
\label{p.25}\end{equation} We may act with a Lorentz boost on the
same particle  by replacing in (\ref{p.23}) the wavepacket $f_0$ by
$f_0^{\Lambda}(x) = f_0(\Lambda^{-1}x)$, $\Lambda \in
L_+^{\uparrow}$; the amplitude is modified as follows
\begin{equation}
\frac 1{\tau (f_0^{\Lambda})} = {(2\pi)\coupl^2\, \int (2(\Lambda
p)^0)^{-1}\,|\wt f_0(p)|^2\,\,\delta(p^2-m_0^2)\theta(p^0)\,dp
\over\int |\wt f_0(p)|^2\,\delta(p^2-m_0^2)\theta(p^0)\,dp}
q!\rho(m_0^2;m_1,\ldots,m_1,\ldots,m_N,\ldots,m_N)\ .
\label{p.23.a}\end{equation} If again $|\wt f_0(p)|^2 \rightarrow
\delta(\vec{p})$, the final result is the expression in (\ref{p.25})
multiplied by $1/(\Lambda)_{00}$. If $\Lambda = \exp(s M_{10})$,
i.e. the particle is moving with velocity ${\rm v}= \th s$,
$(\Lambda)_{00} = \ch s = (1-{\rm v}^2)^{-1/2}$ gives the usual
correction to the lifetime:
\begin{equation}
\tau_{\rm v} = \tau_0/\sqrt {1-{\rm v}^2}. \label{sperel}
\end{equation}
\begin{remark}\rm
It is worthwhile to stress once more that this effect, which expresses
the behavior of the life-time of a moving particle in
special relativity, crucially depends on the peculiar way in which
the wavepacket enters in the transition amplitude per unit time
(\ref{p.23}).
\end{remark}

\subsection{K\"all\'en-Lehmann weights}
The weight $\rho$ can be explicitly computed only in the case of one
particle  decaying into two particles. For a particle of mass $m_0>
0$ decaying into two identical particles of mass $m_1
> 0$, i.e. the case $N=1$, $q_1=2$,
the well-known formula is
\begin{equation}
\rho(m_0^2;\ m_1,m_1) = {\left( {m_0^2 -4m_1^2} \right )^{d-3 \over
2} \over \left(4\pi \right )^{d-1 \over 2} 2^{d-2}\, \Gamma\left
({d-1\over 2}\right )\,m_0 } \ \theta(m_0^2 -4m_1^2),
\label{p.26}\end{equation}
and (\ref{p.25}) becomes
\begin{equation}
\frac 1{\tau_0} = \hbox{(lifetime of $m_0$)}^{-1} =
{\pi\coupl^2\left( {m_0^2 -4m_1^2} \right )^{d-3 \over 2} \over
\left(4\pi \right )^{d-1 \over 2} 2^{d-3}\, \Gamma\left ({d-1\over
2}\right )\,m_0^2 } \ \theta(m_0^2 -4m_1^2).
\label{p.27}\end{equation}
For $d=4$ this is
\begin{equation}
\frac 1{\tau_0} = {\coupl^2\left( {m_0^2 -4m_1^2} \right )^{1 \over
2} \over 8\pi\,m_0^2}\,\theta(m_0^2 -4m_1^2),
\label{p.28}\end{equation}
in agreement with the computation in e.g.
\cite{Veltman}.

\newpage
\section{de Sitter case}
\subsection{Adiabatic limit in the de Sitter case}

\label{dSadlim} The discussion of the adiabatic limit is more
complicated in the de Sitter case. Taking the adiabatic limit is of
course technically much more involved than in the Minkowski case
(and we will relegate all the technical details to the appendices).
But the really
intricate and maybe perplexing issue is the physical interpretation
of the whole procedure and, even more, of the somewhat surprising
results.

Having in mind the Minkowskian case that we have just discussed, the
first question that should be asked is what is "time" in the de
Sitter universe and what does it means that an interaction lasts for
a certain time. In the Minkowski case we have the solid foundation
of special relativity and a privileged class of frames, the inertial
frames, each of them having an inherent precise notion of time.

In the de Sitter case (and the situation is even worse in a general
curved spacetime) we have no such thing. Instead we have  many
possible coordinate systems, that may or may not cover the whole
manifold, and many possible choices of temporal coordinates that
have no special relation to each other.

For example, the de Sitter universe is the only known spacetime
manifold admitting three different inequivalent choices of cosmic
time so that the de Sitter metric takes the appearance of a,
respectively, closed, flat, or open Friedmann-Robertson-Walker
universe. But there are also other possibilities. The choice of time
coordinate made in 1917 by de Sitter in his original papers
\cite{desitter,desitterbis}  describes a wedge-like region of the de Sitter
manifold as a static spacetime with bifurcate Killing horizons \cite{kayw}.

We choose to proceed heuristically in analogy with the Minkowskian
case. Concretely, we will work out the adiabatic limit using two of
the three possible cosmological coordinate systems, namely the
closed and the flat systems. Starting again from eq. (\ref{p.16}) we
take the cutoff $g$ appearing in there as the indicator (or
characteristic) function of some ``cosmic time-slice'' of thickness
$T$ w.r.t. to the relevant choice of cosmic time.

We will see that in both the closed and flat case the amplitude
diverges linearly in $T$ precisely as in the Minkowskian case.
Therefore, to extract a finite limit we are entitled (and have no
other choice than) to use the Fermi golden rule and compute in the
above two frames the probability per unit time by dividing by $T$;
there is at this point a small difference w.r.t. the flat case: the
amplitude per unit time at finite $T$ depends on $T$. However,
letting $T\rightarrow \infty$ gives a well-defined limit which
exhibits a much more disturbing difference with the Minkowskian case.

\paragraph{Closed FRW model:}

The relevant coordinate system is the following:
\begin{equation}
x(t,\vec u) = \left\{\begin{array}{lcl} x^0 &=& R\,\sh(t/R),\\
{x^i} &=& R\,\ch (t/R)\,\ (-\vec{u}),\ \ \ \vec{u} \in S^{d-1}\ ,
\label{s.1}\end{array}\right.
\end{equation}
(the minus sign at rhs is for further convenience). In this
coordinate system the constant time slices are hyperspheres. These
coordinates have the advantage to globally cover the de Sitter
manifold; they gives to the metric the form of a closed FRW model
with scale factor $a(t)= \ch(t/R)$:
\begin{eqnarray}
ds^2 &=& dt^2- R^2\ch^2(t/R) d\sigma^2(\vec{u})\label{s.2.1}\\
&=& dt^2- R^2\ch^2(t/R) (d\chi^2+\sin^2\chi
(d\theta^2+\sin^2\theta\,d\phi^2))\ \ \ (d=4).
\label{s.2.2}\end{eqnarray}
In (\ref{s.2.1}) $d\sigma^2(\vec{u})$ is
the square line element on $S^{d-1}$ at $\vec{u}$, and (\ref{s.2.2})
includes its expression in Euler angles. In these coordinates, we
choose
\begin{equation}
g(x) = g_T(x) = \theta(t+T/2)\,\theta(T/2-t)\ .
\label{s.3}\end{equation}
\begin{figure}[h]
\begin{center}
\includegraphics[width=12cm]{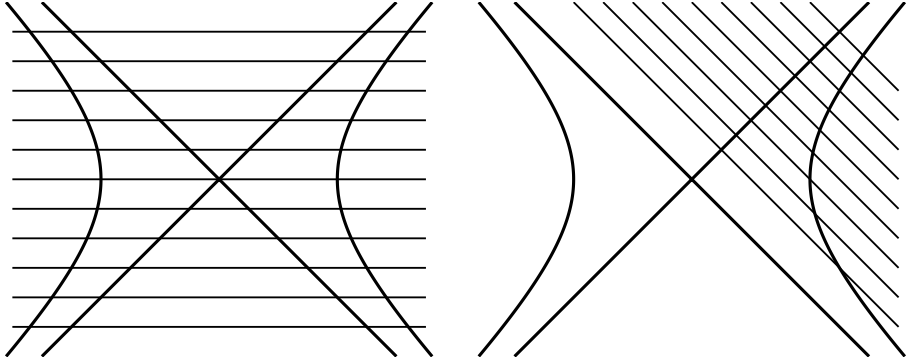}
\caption{Time-slices of the de Sitter spacetime in the closed and in the flat
coordinate systems.} \label{fig:sigma}
\end{center}
\end{figure}
\paragraph{Flat FRW model:}
These are the coordinates currently used in the context of
inflationary models. Hypersurfaces of  constant time are flat:
\begin{equation}
x(t,\y) = \left\{\begin{array}{lcl}
x^{0} &=& R\, \sh  {t\over R} +  {1\over 2R}\, e^{t\over R} \y^2\,,    \\
x^{j} &=&  e^ {t\over R}  \y^j,\ \ (1\le j \le d-1)\,,\\
x^{d} &=& R\, \ch {t\over R} -  {1\over 2R}\,  e^{t\over R} \y^2\,,
\end{array} \right.\ \ \y \in \bR^{d-1},
\label{s.4}\end{equation}
\begin{eqnarray}
ds^2= dt^2-e^{2t/R}(d\y_1^2+\ldots +d\y_{d-1}^2)\ ,
\label{s.5}\end{eqnarray}
 In these coordinates we choose
\begin{equation}
g(x) = g_T(x) = \theta(t+T/2)\,\theta(T/2 -t)\ .
\label{s.6}\end{equation}
But the coordinates (\ref{s.4}) only cover one half of $X_d$, the
region where $x^0+x^d
>0$, and the adiabatic limit will have to include the contribution
of the other half not covered by the coordinate system.

It turns out that the limit

\begin{equation}
\frac {1}{\tau} = \left(\lim _{T\to \infty}\frac{L_1(f_0,\
g)}{T}\right)\times\,
q!\,\rho(m_0^2;m_1,\ldots,m_1,\ldots,m_N,\ldots,m_N)\
\end{equation}
exists and is the same for both kinds of slices. The calculations
are tedious and not quite straightforward, and will be given in
Appendices \ref{adhor} and \ref{adpar}. For the spherical slices of
the closed FRW system, only the calculations for $d=2,\ 3,\ 4$ have been
carried out. The method for the flat FRW coordinates works for all
$d$.

The inverse lifetime that results is
\begin{equation}
\Gamma_{1,q_1,\ldots,q_N}(f_0) = {\coupl^2\pi\coth(\pi\kappa)^2\,R
\over |\kappa|} \times\,
q!\,\rho(m_0^2;m_1,\ldots,m_1,\ldots,m_N,\ldots,m_N)
\label{s.7}\end{equation}
where we denoted $\k = R(m_0^2-m_c^2)^{1/2}$. Note the similarity of
this formula with (\ref{p.25}) and in fact the first factor in
($\ref{s.7}$) tends to the corresponding factor in (\ref{p.25}) when
$R \rightarrow \infty$ at fixed $m_0$. However there is a most
striking difference with (\ref{p.23}): the rhs of ($\ref{s.7}$) does
not depend on $f_0$, the initial wave function of the decaying
particle. Therefore in particular the lifetime of a particle does
not depend on its velocity. We will comment on this feature, at
first sight embarrassing, in the conclusions.

\section{K\"all\'en-Lehmann weights}
As in the flat case, an explicit computation of the
K\"all\'en-Lehmann weight is only possible for decays of one
particle into two. Here  the discussion will be restricted to the
case of a particle of mass $m_0
> 0$ decaying into two particles of equal masses $m_1  = m_2 >0$ and
we suppose at the beginning  $m_1 > m_c$. The more difficult case
$m_1 \not= m_2$ will be treated in a paper in preparation
\cite{BEGMP}. We shall find an explicit $\rho(a^2;m_1,\ m_1)$ such
that
\begin{equation}
\www_{m}(z,\ z')^2 = \int \rho(a^2;m_1,\ m_1)\www_{a}(z,\ z')\,da^2\ .
\label{r.1}\end{equation}
We change to variables $\k = [a^2R^2
-(d-1)^2/4]^{1/2}$, $\nu  = [m_1^2R^2 -(d-1)^2/4]^{1/2}$, and (by
abuse of notation) seek a function $\rho(\k;\nu,\ \nu)$ (mostly
abbreviated as $\rho(\k)$) such that
\begin{equation}
\www_{\nu}(z,\ z')^2 = \int_0^\infty 2\k\,\rho(\k;\nu,\
\nu)\,\www_{\k}(z,\ z')\,d\k\ .
\label{r.2}\end{equation}
By (\ref{f.12}), this is equivalent to
\begin{equation}
C_{d,\nu}^{\prime 2}\,(x^2-1)^{-{d-2\over 4}}
P^{-\frac{d-2}{2}}_{-\frac{1}{2} + i\nu}(x)^2 = \int_0^\infty
2C'_{d,\k}\,\k\,\rho(\k)\, P^{-\frac{d-2}{2}}_{-\frac{1}{2} + i\k}(x)\,d\k \ ,
\label{r.3}\end{equation}
with
\begin{equation}
C'_{d,\nu} = \frac{ \Gamma\left(\frac{d-1}{2} +i \nu\right)
\Gamma\left(\frac{d-1}{2} -i \nu\right)}{2(2\pi)^{\frac{d}{2}} R^{d-2}  }\ .
\label{r.4}\end{equation}
The generalized Mehler-Fock
theorem (\cite{MagnusOS}, p.~398) asserts that
\begin{eqnarray}
g(x) &=& \int_0^\infty P^{\sigma}_{-\frac{1}{2} + i \k} (x)f(\k) d
\k \ \ \ \Longleftrightarrow \cr
f(\k) &=&
\frac{\k}{\pi}\sh(\pi\kappa) \,\Gamma\left(\frac{1}{2}-\sigma
+i\k\right) \Gamma\left(\frac{1}{2}-\sigma -i\k\right) \int_1^\infty
P^{\sigma}_{-\frac{1}{2} + i \k}(x)g(x) dx\ .
\label{r.5}\end{eqnarray}
Therefore (\ref{r.3}) implies
\begin{equation}
\k\,\rho(\k) =
{C_{d,\nu}^{\prime 2}\over 2C'_{d,\k}}\frac{\k}{\pi}\sh(\pi\kappa)
\,\Gamma\left({d-1\over 2} +i\k\right) \Gamma\left({d-1\over 2}
-i\k\right)\times h_d(\k,\ \nu,\ \nu),
\label{r.6}\end{equation}
\begin{equation}
h_d(\k,\ \nu,\ \nu) \bydef \int_1^\infty
(x^2-1)^{-{d-2\over 4}} \left[P^{-\frac{d-2}{2}}_{-\frac{1}{2} + i\nu}(x)\right]^2\,
P^{-\frac{d-2}{2}}_{-\frac{1}{2} + i\k}(x)\, dx\ .
\label{r.7}\end{equation}
It is possible to obtain an explicit
expression of $h_d(\k,\ \nu,\ \nu)$ by using Mellin transform
techniques (see \cite{Marichev}) and a lemma of Barnes (see \cite{Slater}).
Recall that if $\vhi \in \DD((0,\ \infty))$, its Mellin transform
$\wh \vhi$ is given by
\begin{equation}
\wh \vhi(s) = \int_0^\infty \zeta^{s-1} \vhi(\zeta)\,d\zeta\ ,
\label{r.8}\end{equation}
It is entire in $s = \sigma+i\tau$, decreasing faster than any negative
power of $\tau$ for fixed $\sigma$, and
\begin{equation}
\vhi(\zeta) = {1\over 2i\pi}
\int_{\sigma -i\infty}^{\sigma +i\infty} \zeta^{-s} \wh \vhi(s)\,ds\ \ \
\forall \sigma \in \bR,
\label{r.10}\end{equation}
\begin{equation}
\psi(\zeta) = \vhi(1/\zeta)\ \ \ \Longleftrightarrow\ \ \
\wh \psi(s) = \wh \vhi(-s).
\label{r.11}\end{equation}
If $\vhi$, $\vhi_1$, $\vhi_2$ are in $\DD((0,\ \infty))$,
\begin{equation}
\vhi(\zeta) = \int_0^\infty
\vhi_1(\zeta/u)\,\vhi_2(u)\,{du \over u}\ \ \
\Longleftrightarrow\ \ \
\wh \vhi(s) = \wh \vhi_1(s) \wh \vhi_2(s)\ .
\label{r.13}\end{equation}
In particular (Mellin-Plancherel identity)
\begin{equation} \int_0^\infty \vhi_1(u)\,\vhi_2(u)\,{du \over u} =
{1 \over 2i\pi} \int_{\sigma -i\infty}^{\sigma +i\infty} \wh
\vhi_1(-s) \wh \vhi_2(s)\,ds\ .
\label{r.15}\end{equation}
These properties can be extended to other functions and generalized functions
(see \cite{Marichev}), and, in many interesting cases, although the Mellin
transforms are no longer entire, the above formulae
survive provided the integration in (\ref{r.10}) or (\ref{r.15}) is
performed on a suitable contour.

By the change of variable $x = \sqrt{1+\zeta}$ in (\ref{r.7}) we find
\begin{equation}
h_d(\k,\ \nu,\ \nu) = \int_0^\infty G_1(\zeta)\,G_2(\zeta)\,{d\zeta
\over \zeta}, \label{r.19}\end{equation} with
\begin{equation}
G_1(\zeta) = {P_\alpha^\mu(\sqrt{1+\zeta}) \over 2\sqrt{1+\zeta}},\
\ \ G_2(\zeta) = \zeta^{{\mu\over 2}+1}
[P_\beta^\mu(\sqrt{1+\zeta})]^2\ . \label{r.20}\end{equation} and
\begin{equation}
\alpha = -\half +i\k,\ \ \ \beta = -\half +i\nu,\ \ \ \mu =
1-{d\over 2}\ .
\label{r.21}\end{equation}
The Mellin transforms of
$G_1$ and $G_2$ are known (see \cite{Marichev}, 17(1) p. 257, and 28(1) p.
263.)
\begin{eqnarray}
&&\wh G_1(s) = {2^{\mu-1} \over
\Gamma\left ( 1+{\alpha-\mu \over 2} \right )
\Gamma\left ( {1-\mu-\alpha \over 2} \right )}
\Gamma \left [ \begin{array}{l}
s-{\mu\over 2},\ \ 1+{\alpha \over 2} -s,\ \ {1-\alpha\over 2} -s\\
1-{\mu\over 2} -s
\end{array} \right ]\cr
&&\hbox{provided}\ \ \Re \mu < 2\Re s < \min\{2+\Re \alpha,\ 1-\Re
\alpha\}, \label{r.22}\\
&&\wh G_2(s) = {1 \over
\pi^{\half}\Gamma(1+\beta-\mu)\Gamma(-\beta-\mu)}\,\Gamma \left [
\begin{array}{l}
1-{\mu\over 2}+s,\ \ \beta-{\mu\over 2}-s,\ -\beta-1-{\mu\over 2}-s,\
-{1+\mu \over 2}-s\\
-{3\mu\over 2}-s,\ \ -{\mu\over 2}-s
\end{array} \right ]\cr
&&\hbox{provided}\ \ \Re \mu < \Re (s+1-\mu/2) < \min\{-\Re \beta,\
\half\},\ \ \ \mu \notin \bN\ .
\label{r.23}\end{eqnarray}
\begin{remark}\rm
We have actually checked
the above formulae using the methods described in \cite{Marichev}. However
some other formulae appearing in that extremely useful reference
have misprints.
\end{remark}
By the Mellin-Plancherel theorem
\begin{equation}
h_d(\k,\ \nu,\ \nu) = {1 \over 2i\pi}
\int_{\sigma-i\infty}^{\sigma+i\infty} \wh G_1(s)\,\wh G_2(-s)\,ds\ ,
\label{r.24}\end{equation}
and the preceding formulae give
\begin{eqnarray}
&&\wh G_1(s)\,\wh G_2(-s) = {1 \over 2^{d\over 2}\sqrt{\pi}
\Gamma \left [ {d+1\over 4} +{i\k \over 2},\ {d+1\over 4} -{i\k \over 2},\
{d-1\over 2} +i\nu,\ {d-1\over 2} -i\nu \right ]} \times\cr
&& \times \Gamma \left [
\begin{array}{l}
{d-4\over 4}+i\nu+s,\ {d-4\over 4}-i\nu+s,\ {d-4\over 4}+s,\
{3\over 4}+{i\k \over 2}-s,\ {3\over 4}-{i\k \over 2}-s\\
{3d-6 \over 4}+s
\end{array} \right ]\ .
\label{r.25}\end{eqnarray}
It is now possible to use Barnes' Second
Lemma \cite{Slater} p. 112) :

\begin{lemma}[Barnes]
\begin{eqnarray}
&&{1\over 2i\pi} \int_{-i\infty}^{i\infty}
\Gamma \left [
\begin{array}{l}
a_1+s,\ a_2+s,\ a_3+s,\ b_1-s,\ b_2-s\\
c+s
\end{array} \right ] ds =
\nonumber\\
&&=\Gamma \left [
\begin{array}{l}
a_1+b_1,\ a_2+b_1,\ a_3+b_1,\ a_1+b_2,\ a_2+b_2,\ a_3+b_2\\
c-a_1,\ c-a_2,\ c-a_3
\end{array} \right ]\ ,
\label{r.26}\end{eqnarray}
provided
\begin{equation}
a_1+a_2+a_3+b_1+b_2-c = 0 \label{r.27}\end{equation} and that the
contour of integration in (\ref{r.26}) separates the increasing and
decreasing series of poles.
\end{lemma}
As a function of $s$, $\wh G_1(s)\wh G_2(-s)$, as given by (\ref{r.25}),
is, up to a factor, of the form of the integrand of (\ref{r.26}) if
we take
\begin{equation}
a_1= {d-4\over 4}+i\nu,\ a_2= {d-4\over 4}-i\nu,\ a_3= {d-4\over
4},\ b_1= {3\over 4}+{i\k \over 2},\ b_2= {3\over 4}-{i\k \over 2},\
c= {3d-6 \over 4}.
\label{r.28}\end{equation}
This choice satisfies
the condition (\ref{r.27}). Therefore
\begin{eqnarray}
&& h_d(\k,\ \nu,\ \nu) = {1 \over 2^{d\over 2}\sqrt{\pi}
\Gamma \left [ {d+1\over 4} +{i\k \over 2},\ {d+1\over 4} -{i\k \over 2},\
{d-1\over 2} +i\nu,\ {d-1\over 2} -i\nu \right ]} \times
\nonumber\\
&& \times \Gamma \left [
\begin{array}{l}
{d-1\over 4}+{i\k \over 2}+i\nu,\ {d-1\over 4}+{i\k \over 2}-i\nu,\
{d-1\over 4}+{i\k \over 2},\
{d-1\over 4}-{i\k \over 2}+i\nu,\ {d-1\over 4}-{i\k \over 2}-i\nu,\
{d-1\over 4}-{i\k \over 2}\\
{d-1\over 2}-i\nu,\ {d-1\over 2}+i\nu,\ {d-1\over 2}
\end{array} \right ]\ .\hbox to 1cm {\hfill}
\label{r.29}\end{eqnarray}
{}From here on, we will use the notation $\mu = (d-1)/4$.

We can recast the above expression for $h_d(\k,\ \nu,\ \nu)$ by using
Legendre's duplication formula:
\begin{eqnarray}
&& h_d(\k,\ \nu,\ \nu) = {1\over 2^{3-d/2}\pi^{3/2}
\Gamma(2\mu+i\k)\Gamma(2\mu-i\k)
\Gamma(2\mu+i\nu)^2\Gamma(2\mu-i\nu)^2\Gamma(2\mu)} \ \times\cr
&&
\Gamma \left ( \mu+{i\k\over 2}+i\nu \right )
\Gamma \left ( \mu+{i\k\over 2}-i\nu \right )
\Gamma \left ( \mu+{i\k\over 2} \right )^2\ \times\cr
&&
\Gamma \left ( \mu-{i\k\over 2}+i\nu \right )
\Gamma \left ( \mu-{i\k\over 2}-i\nu \right )
\Gamma \left ( \mu-{i\k\over 2} \right )^2\ ,\ \ \ \ \mu={d-1\over 4}\ .
\label{r.29.1}
\end{eqnarray}
In this form the formula is a special case of the formula for
two unequal masses which will appear in \cite{BEGMP}.
We note that in the derivation of (\ref{r.29}) or (\ref{r.29.1})
with $h_d$ defined in (\ref{r.7}), $d$ is not restricted to be an
integer. These formulae hold wherever both sides are defined.
Eqs. (\ref{r.29}) and (\ref{r.6}) give
\begin{eqnarray}
\lefteqn{\k\,\rho(\k;\nu,\ \nu) = {
\k\,\sh(\pi\k)\over
2^{d+2}\pi^{d+3\over 2}R^{d-2}
\Gamma\left ({1\over 2}+\mu+{i\k\over 2} \right )
\Gamma\left ({1\over 2}+\mu-{i\k\over 2} \right )
\Gamma(2\mu)}\ \times}\cr
&&
\Gamma\left (\mu+{i\k \over 2} + i\nu  \right )
\Gamma\left (\mu+{i\k \over 2} - i\nu  \right )
\Gamma\left (\mu+{i\k \over 2}  \right )\ \times\cr
&&
\Gamma\left (\mu-{i\k \over 2} + i\nu  \right ) \Gamma\left
(\mu-{i\k \over 2} - i\nu  \right ) \Gamma\left (\mu-{i\k \over 2}
\right )\ ,
\label{r.30}\end{eqnarray}
or, using $\k\sh(\pi\k) =
\pi[\Gamma(i\k)\Gamma(-i\k)]^{-1}$,
\begin{eqnarray}
\lefteqn{\k\,\rho(\k;\nu,\ \nu) = { 1\over 2^{d+2}\pi^{d+1\over
2}R^{d-2}\,\Gamma(i\k)\Gamma(-i\k) \Gamma\left ({1\over
2}+\mu+{i\k\over 2} \right ) \Gamma\left ({1\over 2}+\mu-{i\k\over
2} \right ) \Gamma(2\mu)}\ \times}\cr && \hbox to 5cm {\hfill}
\prod_{\epsilon = \pm 1} \Gamma\left (\mu+{i\epsilon\k \over 2}
\right ) \prod_{\epsilon' = \pm 1} \Gamma\left (\mu+{i\epsilon\k
\over 2} + i\epsilon'\nu  \right )
\label{r.30.1}\end{eqnarray}
This obviously extends to an even analytic function of $\k$, hence
\begin{equation}
w_\nu(z,\ z')^2 = \int_{-\infty}^\infty
\k\,\rho(\k;\nu,\ \nu)\,w_\k(z,\ z')\,d\k\ .
\label{r.31}\end{equation}
Moreover, for real $\nu$ and $\k\not= 0$,
$\k\,\rho(\k;\nu,\ \nu)$ is strictly positive. This shows that, in
the presence of a suitable interaction term (see (\ref{p.3})), any
``principal'' particle can decay into any pair of equal-mass
``principal'' particles.

\subsection{Minkowskian limit}

Setting $\k = MR>0$ and $\nu = mR>0$ in (\ref{r.30}) gives
\begin{equation}
\rho(MR;mR,\ mR) = {\sh(\pi MR)\over 2^{d+2}\pi^{d+3\over
2}R^{d-2}\Gamma\left ({d-1\over 2}\right )} \left | {\prod_{j=1}^3
\Gamma(x_j+iRu_j) \over \Gamma(x_4+iRu_4)} \right |^2\ ,
\label{u.1}\end{equation}
with
\begin{eqnarray}
&&x_1=x_2=x_3= {d-1\over 4},\ \ x_4 = {d+1\over 4},\cr && u_1 =
{M\over 2}+m,\ \ u_2 = {M\over 2}-m,\ \ u_3 = {M\over 2},\ \ u_4 =
{M\over 2}\ .
\label{u.2}\end{eqnarray}
Recall Stirling's formula
(\cite{bateman1}, p. 47):
\begin{equation}
\Gamma(z) = (2\pi)^\half
e^{-z + (z-\half)\log z}\, \left ( 1+ a_1z^{-1}  + a_2z^{-2} +
O(z^{-3}) \right ),\ \ \ a_1 = 1/12,\ \ a_2 = 1/288\ ,
\label{u.3}\end{equation}
valid for $z \notin \bR_-$.
By a straightforward calculation it follows that if $z=x+iy$ and
$|x|$ remains bounded while $|y|\rightarrow +\infty$,
\begin{equation}
|\Gamma(x+iy)|^2 \sim 2\pi e^{-\pi |y|}\,|y|^{2x-1} \left(1+
{(x-\half)x^2 + 2(a_1x-a_2) + a_1^2\over y^2} \right ).
\label{u.4}\end{equation}
Using this in (\ref{u.1}), we find
that as $R\rightarrow \infty$,
\begin{equation}
R^2 \rho(RM;Rm,\ Rm) \sim {\exp\pi R \left ( {M\over 2}-m - \left
|{M\over 2}-m\right |\right )\over 2^d \pi^{d-1\over 2} \Gamma \left
( {d-1\over 2} \right )M} \left ({M^2-4m^2 \over 4} \right
)^{d-3\over 2}\, (1+AR^{-2})\ ,
\label{u.5}\end{equation}
where
\begin{eqnarray}
A &=& \sum_{j=1}^3 {(x_j-\half)x_j^2 + 2(a_1x_j-a_2) + a_1^2\over
u_j^2} - {(x_4-\half)x_4^2 + 2(a_1x_4-a_2) + a_1^2\over u_4^2}\cr
&=& {17\over 16}\left ( {1\over (M+2m)^2} + {1\over (M-2m)^2} \right
) -{107\over 24 M^2} \ \ \ \hbox{for}\ \ d=4\ .
\label{u.6}\end{eqnarray}
Note that the argument of the exponential
in (\ref{u.5}) is 0 if $M -2m \ge 0$, otherwise $-\pi R(2m-M)$ and,
in this case, $R^2 \rho(RM;Rm,\ Rm)$ tends rapidly to 0. In all
cases, (\ref{u.5}) shows that $R^2 \rho(RM;Rm,\ Rm)$ tends to
$\rho^{\rm Minkowski}(M^2;m,\ m)$ (see (\ref{p.26})).

\subsection{Complementary particles}

One benefit of having the explicit formula (\ref{r.30}) is
being able to examine the case of ``complementary'' particles.
The integrand of (\ref{r.31}) is meromorphic in $\k$ and $\nu$.
We can rewrite (\ref{r.31}) as
\begin{eqnarray}
&&w_\nu(z,\ z')^2 =
\int_{\bR} {\k\,\sh(\pi\k)\over
2^{d+5}\pi^{d+{5\over 2}}\Gamma\left({d\over 2}\right)
\Gamma\left({d-1\over 2}\right)R^{2d-4}}\,
F \left (
{d-1\over 2}+i\k,\ {d-1\over 2}-i\k\ ;\ {d\over 2}\ ;\
{1-\zeta \over 2} \right )\ \times\cr
&&\Gamma \left (\mu+{i\k\over 2} \right)^2
\Gamma \left (\mu-{i\k\over 2} \right)^2\,
\prod_{\epsilon,\ \epsilon' =\pm 1}
\Gamma \left ( \mu +{i\epsilon\k\over 2}+{i\epsilon\nu\over 2} \right )\,
d\k\ ,\ \ \ \ \mu = {d-1\over 4}\ .
\label{c.1}\end{eqnarray}
The integrand is meromorphic in $\k$ and $\nu$. It
has no singularity when both are real. The lhs is holomorphic in
$\{\nu\ :\ \nu \notin \pm i((d-1)/2 + \bZ_+)\}$.
We analytically continue the integral in the variable $\nu$:
choose $\nu$ complex with $\Re \nu >0$ and
$\alpha = \Im \nu >0$. Recall that, for integer $n \ge 0$,
\begin{equation}
z+n \sim 0 \ \ \ \Rightarrow \Gamma(z) \sim {(-1)^n\over n!(z+n)}\ .
\label{c.2}\end{equation}
The poles of the functions $\k \mapsto
\Gamma(\mu \pm i\k/2)$ are at $\k = \pm 2i(\mu+n)$ ($n \ge 0$
integer), and are independent of $\nu$. The other poles of the
integrand are as follows ($n \ge 0$ integer):
\begin{equation}
{i\k \over 2}+\mu \pm i\nu+n \sim 0\ \ \Rightarrow \Gamma \left
({i\k \over 2}+\mu \pm i\nu\right ) \sim {(-1)^{n} \over {i\over
2}n!(\k -2i(\mu \pm i\nu +n))}\ ,
\label{c.3}\end{equation}
\begin{equation}
-{i\k \over 2}+\mu \pm i\nu+n \sim 0\ \ \Rightarrow \Gamma \left
(-{i\k \over 2}+\mu \pm i\nu\right ) \sim {(-1)^{n} \over -{i\over
2}n!(\k +2i(\mu \pm i\nu +n))}\ .
\label{c.4}\end{equation}
The
poles $\k -2i(\mu +i\nu +n) =0$ (see (\ref{c.3})) and the poles $\k
+2i(\mu -i\nu +n) =0$ (see (\ref{c.4})) are on the line $-2\Re \nu
+i\bR$. Their mutual distances do not change as $\nu$ varies, and
they all move down as $\Im\nu$ increases. The poles $\k +2i(\mu
+i\nu +n) =0$ (see (\ref{c.4})) and $\k -2i(\mu -i\nu +n) =0$ (see
(\ref{c.3})) are the opposites of those described before. They lie
on $2\Re \nu +i\bR$ and move up as $\Im\nu$ increases.

If $\Im\nu$ increases from 0 but $0< \Im\nu < \mu$,
no pole reaches the real axis and
the formula (\ref{r.31}) continues to hold.
This is true in particular if $\mu = i\alpha$ with
$0< \alpha< (d-1)/4 = m_c/2$, corresponding to $m_c > m_1 > m_c\sqrt{3}/2$.
If this condition is satisfied and $m_0 > m_c$,
eqs. (\ref{decayy}) and (\ref{p.16}) hold and the adiabatic limit exists
just as in the case $m_1 > m_c$.

When $\Im\nu$ reaches $\mu$ we have
\begin{equation}
w_\nu(z,\ z')^2 = \int_{\CC} \k\,\rho(\k,\ \nu,\ \nu)\,w_\k(z,\
z')\,d\k\ ,
\label{c.5}\end{equation}
where the contour $\CC$ is
obtained from $\bR$ by a small downward excursion to avoid the pole
at $-2\Re\nu$, and another small upward excursion to avoid the pole
at $2\Re\nu$. Once $\mu < \Im \nu < \mu+1$, we can extract the
residues of the poles at $\k = \pm 2i(\mu +i\nu)$. A similar
situation occurs when the successive poles $\k = \pm 2i(\mu
+i\nu+n)$ cross the real axis, so that,
for $\Re\nu >0$, $\Im \nu \ge 0$, $\Im \nu - \mu \notin \bZ$,
$N = \max \left \{j \in \bZ\ :\ j < \Im \nu - \mu\right \}$,
\begin{eqnarray}
\lefteqn{ w_\nu(z,\ z')^2 = \int_\bR \k\,\rho(\k,\ \nu,\
\nu)\,w_\k(z,\ z')\,d\k\ + }\cr && + \sum_{n=0}^N \left
[{A_n(\nu)\over 2}\,w_{2i(\mu +i\nu+n)}(z,\ z') + {A'_n(\nu)\over
2}\,w_{-2i(\mu +i\nu+n)}(z,\ z') \right ]\ .
\label{c.6}\end{eqnarray}
Note that if $\Im \nu < \mu$ ($N < 0$),
the discrete sum is not present. It turns out that $A'_n(\nu) =
A_n(\nu)$, which is consistent with $\k\rho$ being even in $\k$.
Recall that $w_\tau = w_{-\tau}$ for any $\tau$. Thus, for $N = \max
\left \{j \in \bZ\ :\ j < \Im \nu - \mu\right \}$, (always supposing
$\Re\nu>0$),
\begin{equation}
w_\nu(z,\ z')^2 = \int_\bR
\k\,\rho(\k,\ \nu,\ \nu)\,w_\k(z,\ z')\,d\k\ + \sum_{n=0}^{N}
A_n(\nu)\,w_{2i(\mu +i\nu+n)}(z,\ z')\ .
\label{c.7}\end{equation}
We find, for integer $n \ge 0$,
\begin{eqnarray}
\lefteqn{ A_n(\nu) = {(-1)^n \over n!2^{d-1} \pi^{d-1\over2}R^{d-2}
\Gamma(2\mu)} \times} \cr
&&{\Gamma(2\mu+2i\nu+n)\Gamma(-2i\nu-n)\Gamma(2\mu+n)\Gamma(-i\nu-n)
\Gamma(2\mu+i\nu+n) \over \Gamma(-2\mu-2i\nu-2n)
\Gamma(2\mu+2i\nu+2n)\Gamma(\half -i\nu-n)
\Gamma(\half+2\mu+i\nu+n)}\ .
\label{c.8}\end{eqnarray}

If now we let $\nu$ tend to $i\alpha$ ($\Re\nu$ tends to 0),
(\ref{c.7}) will continue to hold provided both parts of the rhs
remain meaningful. Therefore, if $0< \alpha < (d-1)/2$,
$\alpha - \mu \notin \bZ$, and
$N = \max \left \{j \in \bZ\ :\ j < \alpha - \mu\right \}$, $\mu = (d-1)/4$,
\begin{equation}
w_{i\alpha}(z,\ z')^2 = \int_\bR \k\rho(\k;i\alpha,\
i\alpha)\,w_\k(z,\ z')\,d\k + \sum_{n=0}^{N}
A_n(i\alpha)\,w_{2i(\alpha-\mu -n)}(z,\ z')\ .
\label{c.8.1}\end{equation}
\begin{eqnarray}
\lefteqn{ \k\rho(\k;i\alpha,\ i\alpha) = {1\over 2^{d+2} \pi^{1+d
\over 2} R^{d-2}\Gamma(2\mu)}\times} \cr
&& {\Gamma(\mu +{i\k \over 2}-\alpha)\Gamma(\mu -{i\k \over 2}-\alpha)
\Gamma(\mu +{i\k \over 2}+\alpha)\Gamma(\mu -{i\k \over 2}+\alpha)
\Gamma(\mu +{i\k \over 2})\Gamma(\mu -{i\k \over 2})\over
\Gamma(i\k)\Gamma(-i\k)
\Gamma(\mu +{1\over 2}+{i\k \over 2})\Gamma(\mu +{1\over 2}-{i\k \over 2})}
\label{c.9}\end{eqnarray}
This is obviously positive. For
$A_n(i\alpha)$ we find
\begin{eqnarray}
\lefteqn{
A_n(i\alpha) = {1 \over n!2^{d-1} \pi^{d-1\over 2}R^{d-2}\Gamma(2\mu)}
{\Gamma(2\alpha-n)\Gamma(2\mu+n)\Gamma(\alpha-n)\Gamma(2\mu-\alpha+n)\over
\Gamma(2\alpha-2\mu-2n)\Gamma(\half+\alpha-n)\Gamma(\half+2\mu-\alpha+n)}\
\times}\cr
&& \hbox to 8.5cm {\hfill}\times
(-1)^n{\Gamma(2\mu-2\alpha+n)\over \Gamma(2\mu-2\alpha+2n)}
\label{c.10}\end{eqnarray}
The first two factors in this expression
are positive since the arguments of all $\Gamma$ functions are
positive due to $\alpha -\mu- n>0$ and $2\mu -\alpha >0$. The last
factor is of the form
\begin{equation}
{(-1)^n \Gamma(n+x)\over
\Gamma(2n+x)} = (-1)^n \prod_{q}^{2n-1} (q+x)^{-1}\ .
\label{c.11}\end{equation}
The last product contains $n$ negative
factors and the result is positive, so that $A_n(i\alpha)\ge 0$.
Thus the Hilbert space with scalar product given by the lhs of
(\ref{c.8.1}) appears as a direct integral of Hilbert spaces
associated with unitary irreducible representations of $G_0$.

We conclude that

\begin{description}

\item{1.} Any particle from the principal series can decay
into two particles (of equal masses) of any series.

\item{2.} A particle of the complementary series with parameter
$\k = i\beta$, with $0< \beta < 2\mu$ can decay into two
particles with parameter $i\alpha$, $\alpha = \half\beta +\mu +n$,
where $n$ is any integer such that $0 \le n$ and $\alpha < 2\mu$,
i.e. $n < \mu -\beta/2$. This relation can also be written as
\begin{equation}
(2\mu -\beta)  =  2(2\mu - \alpha) + 2n  < 2\mu.
\label{c.12}\end{equation}
This implies a form of particle stability,  but the new phenomenon
is that a particle of this kind cannot disintegrate unless the
masses of the decay products have certain quantized values.
Stability for the same range of masses has also been recently
found  \cite{skenderis} in a completely different context.

\end{description}

\section{Concluding remarks}

In trying to interpret the  results concerning the lack of mass subadditivity in the de Sitter universe, one can wonder
whether they might be due to the thermodynamical properties
(\cite{Gibbons:1977mu,Bros:1994dn,Bros:1995js,Bros:1998ik})
of the fundamental state we have been using.
We have tested this possibility against a similar computation
in flat thermal field theory that however does not exhibit
this phenomenon in two-particle decays.
Another issue has to do with energy conservation and the relation
mass/energy. dS invariant field theories
admit ten conserved quantities (in $d=4$).  The identification
of a conserved energy  among these quantities
has proven to be useful in classical field theory \cite{Abbott:1981ff}.
The same quantity remains exactly
conserved also at the quantum level although it becomes an operator whose spectrum
is not positive \cite{Bros:1994dn,Bros:1995js,Bros:1998ik} even when restricted to the region where the corresponding classical expression is positive \cite{Abbott:1981ff}; the thermodynamical properties of dS fields arise precisely in this restriction \cite{Gibbons:1977mu,Bros:1994dn,Bros:1995js,Bros:1998ik}.
Energy is conserved also in the  decay processes that violate mass subadditivity, once the adiabatic limit has been performed.
The breakdown of the subadditivity property of
masses in dS spacetime just reflects the nonexistence of
an Abelian translation group and thereby
of a linear energy-momentum space.

When we consider the adiabatic limit problem and its meaning in the de Sitter
context a first complication is the existence of several  choices
of cosmic time, having different physical implications and the
result might depend on one's preferred choice. We have studied the closed and the flat cosmological and found that in both models the first factor in
(\ref{decayy}) diverges like $T$; thus it has to be divided by $T$ to
extract a finite result which is the same in both models.

Here the second (unforeseen) result comes in: in contrast to the
Minkowskian case the limiting probability per unit of time  does not
depend on the wavepacket! This result seems to contradict what we
see everyday in laboratory experiments, a well known effect of
special relativity (Eq. \ref{sperel}). Furthermore, in contrast with the
violation of particle stability that is exponentially small in the
de Sitter radius, this phenomenon does not depend on how small is
the cosmological constant. How can we solve this paradox and
reconcile the result with everyday experience? The point is that the
idea of probability per unit time (Fermi's golden rule) has no
scale-invariant meaning in de Sitter: if we use the limiting
probability to evaluate amplitudes of processes that take place in a
short time we get a grossly wrong result. This is in strong
disagreement with what happens in the Minkowski case where the
limiting probability is attained almost immediately (i.e. already
for finite $T$). Therefore to describe what we are really doing in a
laboratory we should not take the limit $T\to \infty$ and rather use
the probability per unit of time relative to a laboratory consistent
scale of time. In that case we will recover all the standard wisdom
even in presence of a cosmological constant. But, if an unstable
particle lives a very long time ($>>R$) and we can accumulate observations
then a nonvanishing cosmological constant would radically
modify the Minkowski result and de Sitter invariant result will
emerge. This result should not be shocking: after all erasing any
inhomogeneity is precisely what the quasi de Sitter phase is
supposed to do at the epoch of inflation; in the same way, from the
viewpoint of an accelerating universe all the long-lived particles
look as if they were at rest and so their lifetime would not depend
on their peculiar motion.

We thank T.~Damour, H.~de~Vega, M.~Gaudin, G.~Gibbons, D.~Marolf,
M.~Milgram and V.~Pasquier for enlightening discussions. U.~M.
thanks the SPhT  and the IHES for hospitality and support.

\newpage
\appendix

\section{Appendix. More details in the Minkowski case}
\label{admink}
In this appendix we study in more detail the adiabatic limit in the
Minkowski case: it is possible to let the two occurences of $g$ in
(\ref{p.13}) tend to 1 together, or even at different rates.
Let
\begin{eqnarray}
\lefteqn{
\UU(f_0,\ \vhi_1,\ \vhi_2,\ \rho) =
\int \ovl{f_0(x)}\,f_0(y)\,\vhi_1(u)\,\vhi_2(v)\,
\rho(\sigma^2)\,}
\cr
&&w_{m_0}(x,\ u)\,w_a(u,\ v)\,w_{m_0}(v,\ y)\,
dx\,du\,dv\,dy\,d\sigma^2 \ .
\label{g.1}
\end{eqnarray}
We will assume that $\rho$ is $\CC^\infty$ and has support in $c^2 + \bR_+$,
with $0< c < m_0$, and that, for each
integer $n \ge 0$, there are constants
$C_n \ge 0$ and $L_n \ge 0$, such that, for all real $t \ge c^2$,
\begin{equation}
|\rho^{(n)}(t)| \le C_n (1+|t|)^{L_n}. \label{g.111.1}\end{equation}
We take
\begin{equation}
\vhi_j(x) = \int_{\bR^d} e^{-ipx}\wt \vhi_j(p)\,dp,\ \ \ j= 1,\ 2,
\label{g.2}\end{equation}
\begin{equation}
\wt \vhi_j(p) = \veps_j^{-1}\wh g_j(p^0/\veps_j)\,\wt \psi(p^0)
\delta(\vec{p}),\ \ \ \vhi_j(x) = {1\over 2\pi}\int_\bR g_j(\veps_j
(x^0 -t))\psi(t)\,dt, \ \ \ (j = 1,\ 2), \label{g.7}\end{equation}
\begin{equation}
g_j(t) = \int_\bR e^{-itw} \wh g_j(w)\,dw,\ \ \ \ \psi(t) = \int_\bR
e^{-itw} \wt \psi(w)\,dw. \label{g.7.1}\end{equation} Here $\veps_j
= T_j^{-1} >0$. The function $\psi$ belongs to $\SS(\bR)$ with $\wt
\psi(0)= 1$. The function $g_j$ is $L^\infty$ with compact
support.(The cases of real interest are $g_j(t) = \theta(1/2-|t|)$
or $g_j(t) = \theta(t)\theta(1-t)$.) We find, after using the
various delta-functions,
\begin{eqnarray}
\lefteqn{ \UU(f_0,\ \vhi_1,\ \vhi_2,\ \rho) =}\cr &&(2\pi)^{d+3}
\veps_1^{-1} \veps_2^{-1} \int_{p \in \bR^d\atop h^0 \in \bR}
(2p^0)^{-1} |\wt f_0(p)|^2\, \wh g_1 \left ( {p^0-h^0 \over \veps_1}
\right ) \wh g_2 \left ( {h^0-p^0 \over \veps_2} \right ) \wt
\psi(p^0-h^0) \wt \psi(h^0-p^0) \cr &&\delta(p^2-m_0^2)\theta(p^0)\,
\rho(h_0^2-p_0^2 + m_0^2)\theta(h^0)\,dp\,dh^0\ .
\label{g.10}\end{eqnarray}
We now change from the variable $h^0$ to
the variable $w$ such that $h^0 = p^0 +w$\ :
\begin{equation}
\veps_2 \UU(f_0,\ \vhi_1,\ \vhi_2,\ \rho) = \veps_1^{-1}\int \wt
H(w)\,\wh g_1 \left (-{w\over \veps_1} \right ) \wh g_2 \left ({w
\over \veps_2} \right )\,dw = \int \wt H(\veps_1 r)\, \wh g_1(-r)
\wh g_2 \left ({\veps_1 r \over \veps_2} \right )\,dr.
\label{g.11.1}\end{equation}
Here
\begin{equation}
\wt H(w) = (2\pi)^{d+3} |\wt \psi(w)|^2 \int (2p^0)^{-1} |\wt
f_0(p)|^2\, \delta(p^2-m_0^2)\theta(p^0)\,\rho(m_0^2 + w(2p^0+w))\,
\theta(p^0+w)\,dp\ ,
\label{g.11.2}\end{equation}
and we set
\begin{equation}
H(t) = \int_\bR e^{-itw} \wt H(w)\,dw\ .
\label{g.11.3}\end{equation}
Then
\begin{equation}
\veps_2 \UU(f_0,\ \vhi_1,\ \vhi_2,\ \rho) = (2\pi)^{-2} \int_{\bR^2}
H(x)\,g_1(\veps_1 x+ {\veps_1\over \veps_2}y)\,g_2(y)\,dx\,dy\ .
\label{g.11.4}\end{equation}
With our assumptions on $\rho$, $H\in
\SS(\bR)$. Since $g_j$ is $L^\infty$ with compact support and $H \in
\SS(\bR)$, the above integral (\ref{g.11.4}) is absolutely
convergent, uniformly in $\veps_1$ and $\veps_2$. Hence
\begin{equation}
\veps_2 \UU(f_0,\ \vhi_1,\ \vhi_2,\ \rho) = (2\pi)^{-2} \int_{\bR}
H(x)\,G(x,\ \veps_1,\ \veps_2)\,dx,
\label{g.11.5}\end{equation}
\begin{equation}
G(x,\ \veps_1,\ \veps_2) = \int_\bR g_1 \left (\veps_1 x+
{\veps_1\over \veps_2}y \right )\,g_2(y)\,dy = \int_\bR g_1\left (
{\veps_1\over \veps_2}y \right )\, g_2(y- \veps_2 x)\,dy\ .
\label{g.11.6}\end{equation}
We assume from now on $0 < \veps_1 \le
\veps_2 \le 1$. Since $g_j \in L^\infty\cap L^1$ and translation is
continuous on $L^1$, $G$ is continuous in $x$.

\noindent {\bf Example 1.} We suppose that $g_1$, $g_2$ are $\CC^\infty$
with compact support. In this case the limits when $\veps_j$ tend to 0
can be taken under the integral sign in (\ref{g.11.6}).
\begin{description}
\item{(1.1)} if $\veps_1$ tends to 0 at fixed $\veps_2$,
$G$ tends to the constant $g_1(0)\int g_2(y)\,dy$, independent of $\veps_2$.
\item{(1.2)} if both $\veps_1$ and $\veps_2$ tend to 0 and
$\veps_1/\veps_2 \rightarrow 0$, $G$ also tends to $g_1(0)\int g_2(y)\,dy$.
\item{(1.3)} if both $\veps_1$ and $\veps_2$ tend to 0 and
$\veps_1/\veps_2 \rightarrow \lambda \in (0,\ 1]$, then
$G$ tends to the constant $\int g_1(\lambda y)g_2(y)\,dy$, and
\begin{equation}
\veps_2 \UU(f_0,\ \vhi_1,\ \vhi_2,\ \rho) \rightarrow
(2\pi)^{-2}\int H(x)\,dx \int g_1(\lambda y)g_2(y)\,dy.
\label{g.11.6.1}\end{equation}
This holds in particular if $\veps_1$
and $\veps_2$ are kept equal so that $\lambda =1$. The constant
$\int g_1(\lambda y)g_2(y)\,dy$ may be equal to the preceding
constant $g_1(0)\int g_2(y)\,dy$, for example if $g_1(\lambda y) =
g_1(0)$ on the support of $g_2$.
\end{description}

\noindent {\bf Example 2.}
We consider the case when
$g_j(x) = \theta(x)\theta(1-x)$, i.e. $g_j$ is the indicator function
of $[0,\ 1]$. Then (see Fig. \ref{fig1})
\begin{eqnarray}
G(x,\ \veps_1,\ \veps_2) &=& (1+\veps_2 x)\theta(1+\veps_2
x)\theta(-x) + \theta(x)\theta(\veps_1^{-1} - \veps_2^{-1} -x) +\cr
&+& \veps_2(\veps_1^{-1} - x)\theta(x - \veps_1^{-1} + \veps_2^{-1})
\theta(\veps_1^{-1} - x)\ .
\label{g.11.8}\end{eqnarray}
\setlength{\unitlength}{1 mm}
\begin{figure}[h]
\begin{picture}(160,40)(-80,-10)
\put(-80,0){\line(1,0){160}}
\put(-70,0){\line(4,1){40}}
\put(-30,10){\line(1,0){60}}
\put(30,10){\line(4,-1){40}}
\put(-30,-10){\line(0,1){30}}
\put(30,0){\line(0,1){10}}
\put(-75,-5){$\veps_2^{-1}$}
\put(-35, -5){$0$}
\put(25, -5){$\veps_1^{-1} - \veps_2^{-1}$}
\put(65,-5){$\veps_1^{-1}$}
\put(-35,12){$1$}
\end{picture}
\caption{\label{fig1}Graph of $G(x,\ \veps_1,\ \veps_2)$ when
$g_1(x) = g_2(x) = \theta(x)\theta(1-x)$.}
\end{figure}
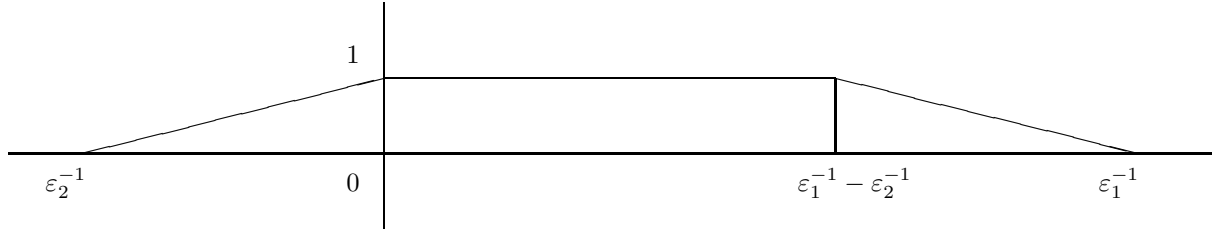
$G(x,\ \veps_1,\ \veps_2)$ tends to 1 when both
$\veps_j \rightarrow 0$ (with $\veps_1 \le \veps_2$).
$G$ also tends to 1 if $\veps_1 \rightarrow 0$ at fixed $\veps_2$
and then $\veps_2 \rightarrow 0$.
Since $H \in \SS(\bR)$, the integral (\ref{g.11.5}) tends to
$(2\pi)^{-2} \int_{\bR} H(x)\,dx = (2\pi)^{-1}\wt H(0).$

\begin{figure}[h]
\begin{picture}(160,40)(-80,-10)
\put(-80,0){\line(1,0){160}}
\put(0,-10){\line(0,1){30}}
\put(-70,0){\line(4,1){40}}
\put(-30,10){\line(1,0){60}}
\put(30,10){\line(4,-1){40}}
\put(-30,0){\line(0,1){10}}
\put(30,0){\line(0,1){10}}
\put(-75,-5){$-{1\over 2}(\veps_1^{-1} + \veps_2^{-1})$}
\put(-42, -5){$-{1\over 2}(\veps_1^{-1} - \veps_2^{-1})$}
\put(25, -5){${1\over 2}(\veps_1^{-1} - \veps_2^{-1})$}
\put(60,-5){${1\over 2}(\veps_1^{-1} + \veps_2^{-1})$}
\put(-5,-5){0}
\put(-5,12){$1$}
\end{picture}
\caption{\label{fig2}Graph of $G(x,\ \veps_1,\ \veps_2)$ when
$g_1(x) = g_2(x) = \theta(1/2-|x|)$.}
\end{figure}
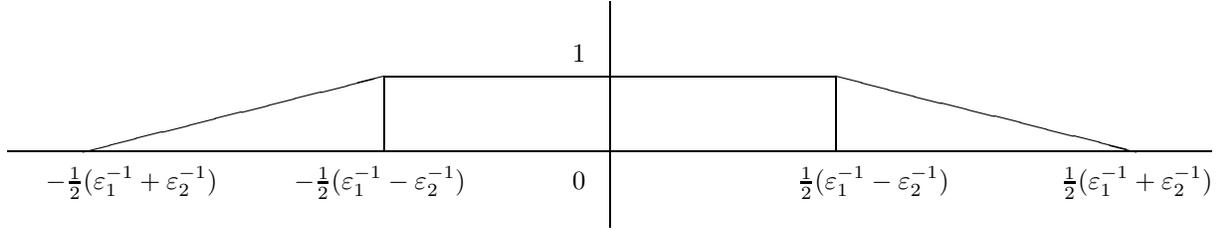

\noindent {\bf Example 3.}
Consider now the case $g_j(x) = \theta(1/2 - |x|)$, i.e.
$g_j$ is the indicator function of $[-1/2,\ 1/2]$.
In that case $G(x,\ \veps_1,\ \veps_2)$ is even in $x$ (see Fig. \ref{fig2}):
\begin{eqnarray}
G(x,\ \veps_1,\ \veps_2) &=& \theta(\veps_1^{-1} - \veps_2^{-1}
-2|x|)\cr &+& \left ( {1\over 2}\left ( 1+ {\veps_2\over \veps_1}
\right ) -\veps_2 |x| \right ) \theta(2|x| - \veps_1^{-1} +
\veps_2^{-1}) \theta(\veps_1^{-1} + \veps_2^{-1} -2|x|).
\label{g.11.9}\end{eqnarray}

$G(x,\ \veps_1,\ \veps_2)$ tends to 1 either if $\veps_1$ tends to
0 at fixed $\veps_2$, or if both
$\veps_j \rightarrow 0$ (with $\veps_1 \le \veps_2$), and
the integral (\ref{g.11.5}) tends to
$(2\pi)^{-2} \int_{\bR} H(x)\,dx = (2\pi)^{-1}\wt H(0)$.

\noindent{\bf Conclusion}
With the two last choices of $g_j$ just described,
\begin{equation}
\veps_2 \UU(f_0,\ \vhi_1,\ \vhi_2,\ \rho) \rightarrow
(2\pi)^{d+2}\,\rho(m_0^2) \int (2p^0)^{-1}|\wt f_0(p)|^2
\delta(p^2-m_0^2)\theta(p^0)\,dp\ .
\label{g.11.9.1}\end{equation}
For other choices of $g_j$, the limit as $\veps_1 \rightarrow 0$,
then $\veps_2 \rightarrow 0$ need not be the same as when $\veps_1 =
\veps_2 \rightarrow 0$.

If we relax the conditions set on $\rho$, the same conclusions
hold if e.g. $\rho(s) = \rho(s;m_1,m_2)$ and $d \ge 4$.

\section{Appendix. Adiabatic limit (dS): horizontal slices}

\label{adhor}
Horizontal slices have been described in subsect. \ref{dSadlim}.
In this appendix, we study $\lim_{T \rightarrow +\infty} T^{-1}L_1(f_0,\ g)$
where $g(x)$ is given by (\ref{s.3}) in the coordinates (\ref{s.1}).
We denote $\k = [m_0^2-(d-1)^2/4]^{1/2}$ and recall that $m_0 > (d-1)/2$
(hence $\k > 0$),
and $C_1(m_o,\ d) = C_0(\k)$ (see (\ref{f.19}).
Inserting the representation (\ref{f.16.2}) for the three occurrences
of $w_{m_0+}$ (denoted also $w_{\k+}$)
in the formula (\ref{p.16}) for $L_1(f_0,\ g)$ gives
\begin{eqnarray}
&& L_1(f_0,\ g) =
{\coupl^2\,C_0(\k)\,c_{d,\k} \int_{\gamma\times\gamma}\ovl{h_0(\xi)}\,
K_\k(\xi,\xi',g)\,h_0(\xi')\,d\mu_\gamma(\xi)\,d\mu_\gamma(\xi') \over
\int_{\gamma} \ovl{h_0(\xi)}\,h_0(\xi)\,d\mu_\gamma(\xi)}\ ,
\label{ah.1}\\
&&h_0(\xi) = \int_{X_d}\psi_{-{d-1\over 2}-i\k}^+(x,\ \xi)\,f_0(x)\,dx\ ,
\label{ah.2}\\
&& K_\k(\xi,\xi',g) = \int_{X_d} \,\psi_{-{d-1\over 2} -i\k}^{+} (x
,\xi)\, \psi_{-{d-1\over 2} +i\k}^{-} (x ,\xi')\,g(x)\,dx\ .
\label{ah.3}\end{eqnarray} We take $\gamma = S_0 = \{\xi \in C_+\ :\
\xi^0 =1\} \simeq S^{d-1}$, the unit sphere in $\bR^d$. In this
appendix, we also set $R=1$: a general $R$ can be reinstated in the
results by homogeneity. Note that, for any $\vhi \in
\CC^\infty(S_0\times S_0)$, $\int \vhi(\xi,\ \xi')\,\psi_{-{d-1\over
2} -i\k}^{+} (x ,\xi)\, \psi_{-{d-1\over 2} +i\k}^{-} (x
,\xi')\,d\mu_{S_0}(\xi)\,d\mu_{S_0}(\xi')$ is $\CC^\infty$ in $x$.
Hence, for any bounded $g$ with bounded support, $K_\k$ is a
distribution on $S_0\times S_0$ in the variables $(\xi,\ \xi')$. We
will take $g$ invariant under the rotation group in $d$ dimensions
(leaving $e_0$ invariant), hence $K_\k(\xi,\xi',g) =
K_\k(L\xi,L\xi',g)$ for every such rotation $L$. Hence $K_\k$ is
$\CC^\infty$ in $\xi$ when smeared with a test-function in $\xi'$.
Studying the limit of $T^{-1}L_1(f_0,\ g)$ for $g$ as in
(\ref{s.3}), is therefore equivalent to studying the limit of
$T^{-1}K_\k(\xi,\xi',g)$ as a distribution in $\xi'$ for fixed
$\xi$. In this appendix the cases $d=2$ and $d=4$ will be treated.
The case $d=3$, more straightforward than $d=4$ (no need to use
$d=2$), will be omitted. The result in these three cases is the same
(see (\ref{ah.31}) and (\ref{n4.20})).

\subsection{Case $d=2$}

We use the following parametrizations
\begin{equation}
\left \{
\begin{array}{l}
x^0 = \sh t\\
x^1 = \ch t\,\sin \theta\\
x^2 = \ch t\,\cos \theta
\end{array}\right .\ \ \ 
\left \{
\begin{array}{l}
\xi^0 = 1\\
\xi^1 = 0\\
\xi^2 = -1
\end{array}\right .\ \ \
\left \{
\begin{array}{l}
\xi'^0 = 1\\
\xi'^1 = -\sin \phi\\
\xi'^2 = -\cos \phi
\end{array}\right .
\label{ah.4}\end{equation} with
\begin{equation}
t \in \bR,\ -\pi < \theta < \pi,\ \ (x \in X_2),\ \ -\pi < \phi <
\pi,\ \ (\xi,\ \xi' \in \partial V_+). \label{ah.5}\end{equation} In
these variables, the measure $dx$ takes the form $\ch
t\,dt\,d\theta$. For small $\veps >0$, changing $t$ into $t\pm
i\veps$ pushes $x$ into $\TT_\pm$. If $g(x)=g_0(t)$,
\begin{eqnarray}
\lefteqn{
K_\k(\xi,\xi',g)  = \int_{t \in \bR,\ -\pi\le\theta\le\pi}g_0(t)}
\nonumber\\
&&
[\sh(t+ i0) + \ch(t+ i0) \cos(\theta)]^{-\half - i\k}
[\sh(t- i0) + \ch(t- i0) \cos(\theta -\phi)]^{-\half + i\k}\,
\ch t\,dt\,d\theta .\hbox to 1 cm{\hfill}
\label{ah.6}\end{eqnarray}
For real $s$ with $0< |s|<\pi/2$ and real $\alpha$,
\begin{equation}
\sh(t+ is) + \ch(t+ is) \cos(\alpha) = \cos(s)(\sh t +\ch t
\cos(\alpha)) + i\sin(s)(\ch t + \sh t\cos(\alpha))
\label{ah.6.1}\end{equation} has a non-zero imaginary part of the
same sign as $s$, so its power $\mu$ can be taken for any complex
$\mu$ and remains analytic in $t$ for all real $t$, smooth and
periodic with period $2\pi$ in $\alpha$. The integral over $\theta$
in (\ref{ah.6}) will be performed, using Plancherel's formula, by
first computing the discrete Fourier transform, in the variable
$\theta$, of the two last factors in the integrand, i.e.
\begin{equation}
K_\k(\xi,\xi',g)  = {1\over 2\pi} \sum_{m \in \bZ} \int_{t \in \bR}
e^{im\phi} F_m(t+i0)\,\ovl{F_m(t+i0)}\,g_0(t)\,\ch(t)dt,
\label{ah.7}\end{equation}
with
\begin{equation}
F_m(t+is) = \int_{-\pi}^{\pi} [\sh(t+ is) + \ch(t+ is)
\cos(\theta)]^{-\half - i\k}\, e^{im\theta}d\theta\ .\ \ \ (0 <|s| <
\pi/2). \label{ah.8}\end{equation}
By changing $\theta$ to $-\theta$
in the integration, we get:
\begin{equation}
F_m(t+is) = F_{-m}(t+is).
\label{ah.8.1}\end{equation}
We use the
formula (\cite{bateman1}, (15) p.157)
\begin{equation}
P_\mu^m(z) = {\Gamma(\mu +m +1) \over 2\pi \Gamma(\mu+1)}
\int_{-\pi}^{\pi} [z + (z^2-1)^{1/2}\,\cos \phi]^\mu\,
e^{im\phi}d\phi\ ,\ \ \ (z \in \Delta_1,\ \ \Re z >0).
\label{ah.9}\end{equation}
This is stated for $m$ integer and $\ge
0$, but by using (see \cite{bateman1}, (7) p.140)
\begin{equation}
{P_\lambda^{-m}(z)\over \Gamma(\lambda-m+1)}
 = {P_\lambda^{m}(z)\over \Gamma(\lambda+m+1)},
\ \ \ m\in \bZ\ ,
\label{ah.10}\end{equation}
it is seen to hold for
all $m\in \bZ$. This gives, for $0< s <\pi/2$,
\begin{eqnarray}
F_m(t+is) &=& e^{-i\pi/4 +\pi\k/2}\,
{2\pi \Gamma(\half -i\k) \over \Gamma(m+\half -i\k)}
P_{-\half -i\k}^m(-i\sh(t+is))\ \ \ {\rm for}\ t>0,\label{ah.11}\\
F_m(t+is) &=& (-1)^m e^{-i\pi/4 +\pi\k/2}\,
{2\pi \Gamma(\half -i\k) \over \Gamma(m+\half -i\k)}
P_{-\half -i\k}^m(-i\sh(t+is))\ \ \ {\rm for}\ t<0,
\label{ah.12}\end{eqnarray}
\begin{eqnarray}
\ovl{F_m(t+is)} &=&
e^{i\pi/4 +\pi\k/2}\,
{2\pi \Gamma(\half +i\k) \over \Gamma(-m+\half +i\k)}
P_{-\half +i\k}^{-m}(i\sh(t-is))\ \ \ {\rm for}\ t>0,\label{ah.13}\\
&=& (-1)^m e^{i\pi/4 +\pi\k/2}\, {2\pi \Gamma(\half +i\k) \over
\Gamma(-m+\half +i\k)} P_{-\half +i\k}^{-m}(i\sh(t-is))\ \ \ {\rm
for}\ t<0.
\label{ah.14}\end{eqnarray}
Taking $g_0(t) = \theta(T/2
-t)\theta(t+T/2)$, we rewrite (\ref{ah.7}) as \begin{equation}
K_\k(\xi,\xi',g) = \sum_{m \in \bZ} e^{im\phi}\, (I_m^+ + I_m^-),
\label{ah.15}\end{equation}
where
\begin{eqnarray}
I_m^+ &=& {1\over 2\pi} \int_0^{T/2}
F_m(t+i0)\,\ovl{F_m(t+i0)}\,\ch(t)dt
\label{ah.16}\\
&=& 2 \pi e^{\pi \k}(-1)^m\, \int_0^{\sh(T/2)} P_{-\half
-i\k}^m(-iu+\veps)\, P_{-\half +i\k}^{-m}(iu+\veps)\,du.
\label{ah.17}\end{eqnarray} In the last expression we have used
$\Gamma(\half +z)\Gamma(\half -z) = \pi/\cos(\pi z)$, and changed to
the variable $u=\sh t$. Similarly (with now $u = -\sh t$),
\begin{eqnarray}
I_m^- &=& {1\over 2\pi} \int_{-T/2}^0
F_m(t+i0)\,\ovl{F_m(t+i0)}\,\ch(t)\,dt
\label{ah.18}\\
&=& 2 \pi e^{\pi \k}(-1)^m\, \int_0^{\sh(T/2)} P_{-\half
-i\k}^m(iu+\veps)\, P_{-\half +i\k}^{-m}(-iu+\veps)\,du.
\label{ah.19}\end{eqnarray}
\begin{remark}
\label{k}\rm
Using ({ah.10}) and $\Gamma(\half +z)\Gamma(\half -z) = \pi/\cos(\pi z)$
shows that the rhs of (\ref{ah.19}) can be obtained
from the rhs of (\ref{ah.17})
by changing {\it in the integrand} (but not outside the integral)
$\k$ to $-\k$. Note also that $I_m^{\pm} = I_{-m}^{\pm}$ by (\ref{ah.8.1}).
\end{remark}

The meaning of (\ref{ah.15}) is that $K_\k(\xi,\xi',g)$ is a distribution
in $\xi'$ as expressed in the coordinate $\phi$, and that
$m \mapsto I_m^+ + I_m^-$ is its discrete Fourier transform.
It is tempered, i.e. $|I_m^+ + I_m^-|$ does not increase faster than
a power of $|m|$ as $|m| \rightarrow \infty$. To prove
that $T^{-1}K_\k$ tends to a limit (also a distribution in $\phi$)
as $T \rightarrow \infty$ is equivalent to proving that

\begin{description}

\item{(1)}
For each $m$, $T^{-1}(I_m^+ + I_m^-)$ tends to a limit $U_m$ as
$T \rightarrow \infty$,

\item{(2)}
there are two positive constants $P$ and $Q$ such that
$T^{-1}|I_m^+ + I_m^-| \le P(1+|m|^Q)$ for all $m$ and $T$.

\end{description}

\noindent
If both conditions are satisfied, $U_m$ is the $m^{\rm th}$ Fourier
coefficient of the limit, i.e.
$\lim T^{-1}K_\k = \sum_m U_m e^{im\phi}$.

\subsubsection{Condition (1)}
\label{ad2}

We need the asymptotic behavior of $P_\lambda^m(z)$ as $|z|\rightarrow \infty$,
as described in \cite{bateman1}, pp. 123, 124, 126, and 164.
For $z \in  \Delta_1$ and $\zeta = z^{-2}$,
\begin{eqnarray}
P_\lambda^\mu(z) &=&
{2^{-\lambda-1}\pi^{-1/2} \Gamma(-\half-\lambda)\,
z^{-\lambda-1+\mu}\,(z^2-1)^{-\mu/2} \over \Gamma(-\lambda-\mu)}
F(\half +\lambda/2 -\mu/2,\ 1 +\lambda/2 -\mu/2;\ \lambda +3/2;\ \zeta)
\nonumber\\
&+& {2^{\lambda}\pi^{-1/2} \Gamma(\half+\lambda)\,
z^{\lambda+\mu}\,(z^2-1)^{-\mu/2} \over \Gamma(1+\lambda -\mu)}
F(-\lambda/2 -\mu/2,\ \half-\lambda/2 -\mu/2;\ \half -\lambda;\ \zeta).
\label{ah.20}
\end{eqnarray}
If $\lambda +\half \notin \bZ$, the two hypergeometric functions
can be expanded into convergent power series for $|\zeta|<1$.
For $z \in \Delta_1$ and $|z| \rightarrow \infty$, we find
\begin{equation}
P_\lambda^\mu(z) \sim {2^{-\lambda-1}\pi^{-1/2}
\Gamma(-\half-\lambda)\,z^{-\lambda-1} \over \Gamma(-\lambda-\mu)} +
{2^{\lambda}\pi^{-1/2} \Gamma(\half+\lambda)\,z^{\lambda} \over
\Gamma(1+\lambda -\mu)} \label{ah.21}\end{equation} Hence, as $u
\rightarrow +\infty$,
\begin{eqnarray}
\lefteqn{
P_{-\half -i\k}^m(-iu)\,P_{-\half +i\k}^{-m}(iu) \sim}\nonumber\\
&&\left [
{2^{-\half+i\k} \pi^{-1/2} \Gamma(i\k)\,e^{i\pi/4 +\pi\k/2}\,
u^{-\half+i\k} \over \Gamma(\half+i\k-m)}
+ {2^{-\half-i\k}\pi^{-1/2} \Gamma(-i\k)\,e^{i\pi/4 -\pi\k/2}\,
u^{-\half-i\k} \over \Gamma(\half-i\k-m)}\right ]\times\nonumber\\
&& \left [ {2^{-\half-i\k} \pi^{-1/2} \Gamma(-i\k)\,e^{-i\pi/4
+\pi\k/2}\, u^{-\half-i\k} \over \Gamma(\half-i\k+m)} +
{2^{-\half+i\k}\pi^{-1/2} \Gamma(i\k)\,e^{-i\pi/4 -\pi\k/2}\,
u^{-\half+i\k} \over \Gamma(\half+i\k+m)}\right ].
\label{ah.22}\end{eqnarray} We first consider the `off-diagonal
terms' of this product: \begin{equation} {2^{-1+2i\k}
\pi^{-1}\Gamma(i\k)^2\, u^{-1+2i\k} \over
\Gamma(\half+i\k-m)\Gamma(\half+i\k+m)} + {2^{-1-2i\k}
\pi^{-1}\Gamma(-i\k)^2\, u^{-1-2i\k} \over
\Gamma(\half-i\k-m)\Gamma(\half-i\k+m)}\ .
\label{ah.23}\end{equation} These two terms are exchanged by
changing $\k$ to $-\k$. The contribution of the first to $I_m^+/T$
is of the form \begin{equation} {\rm Const.}\, {1\over
T}\int_1^{\sh(T/2)} u^{-1+2i\k}du = {\rm Const.}\, {1\over 2i\k
T}(\sh(T/2)^{2i\k} -1). \label{ah.24}\end{equation} This tends to
zero as $T \rightarrow +\infty$. The same happens for the second
term. The `diagonal terms' are \begin{equation}
{2^{-1}\pi^{-1}\Gamma(i\k)\Gamma(-i\k)\,e^{\pi\k}\,u^{-1} \over
\Gamma(\half+i\k-m)\Gamma(\half-i\k+m)} +
{2^{-1}\pi^{-1}\Gamma(-i\k)\Gamma(+i\k)\,e^{-\pi\k}\,u^{-1} \over
\Gamma(\half-i\k-m)\Gamma(\half+i\k+m)}\ .
\label{ah.25}\end{equation} Again these two terms are exchanged by
changing $\k$ to $-\k$. Their sum can be reexpressed as
\begin{equation} {(-1)^m \ch(\pi\k)^2\,u^{-1} \over
\pi\k\sh(\pi\k)}\ . \label{ah.26}\end{equation} Since
\begin{equation} \int_1^{\sh(T/2)} u^{-1}du = \log(\sh(T/2)) \sim
T/2, \label{ah.27}\end{equation}
\begin{equation} {1\over T} I_m^+ \sim {e^{\pi\k}\ch(\pi\k)^2 \over
\pi\k\sh(\pi\k)}\ . \label{ah.28}\end{equation} Because of Remark
\ref{k}, $I_m^-/T$ has the same limit as $I_m^+/T$ and
\begin{equation} U_m = \lim_{T \rightarrow +\infty}{1 \over T}
(I_m^+ + I_m^-) = {2e^{\pi \k}\ch(\pi\k)^2 \over \k\,\sh(\pi\k)}\ .
\label{ah.29}\end{equation} $U_m$ is independent of $m$, so that if
Condition (2) is satisfied,
\begin{equation} \lim_{T \rightarrow +\infty} {1\over
T}\,K_\k(\xi,\xi',g_T)  = {4\pi\,e^{\pi \k}\ch(\pi\k)^2 \over
\k\,\sh(\pi\k)}\, \delta(\phi) = {4\pi\,e^{\pi \k}\ch(\pi\k)^2 \over
\k\,\sh(\pi\k)}\, \delta_{S^1}(\xi,\ \xi')\ ,
\label{ah.30}\end{equation} and (see (\ref{ah.1}) \begin{equation}
\lim_{T \rightarrow +\infty} T^{-1}\,L_1(f_0,\ g_T) =
\coupl^2\,C_0(\k)\,c_{2,\k} {4\pi\,e^{\pi \k}\ch(\pi\k)^2 \over
\k\,\sh(\pi\k)} = {\coupl^2\,\pi\,\coth(\pi\k)^2\over |\k|}.
\label{ah.31}\end{equation} We note that, owing to the delta
function in (\ref{ah.30}) the dependence on $h_0$  (i.e. on $f_0$)
has completely disappeared from the limit. This result agrees with
(\ref{s.7}).

\subsubsection{Condition (2)}
\label{cond2}

In this subsubsection $\lambda$ always denotes $-\half -i\k$ with
$\k \in \bR$ and $\k \not= 0$.
We first return to the first step of the preceding subsection
in the case $m=0$.
{}From the identity (\ref{ah.20}) and the analyticity of
$\zeta \mapsto F(a,\ b,\ c;\zeta)$ in the unit disk,
it follows that there is a $M_0(\k) >0$ such that
\begin{equation}
\left | P_\lambda (z) - {2^{-\half+i\k} \pi^{-1/2}
\Gamma(i\k)\,z^{-\half+i\k} \over \Gamma(\half+i\k)} -
{2^{-\half-i\k}\pi^{-1/2} \Gamma(-i\k)\,z^{-\half-i\k} \over
\Gamma(\half -i\k)} \right | < M_0(\k)\,|z|^{-5/2}\ \ \ z \in
\Delta_1,\ \ |z|>2. \label{ah.32}\end{equation} By
(\ref{ah.11}-\ref{ah.12}), there is also an $M_1(\k) > 0$ such that,
for $0< s< \pi/2$,
\begin{equation}
|F_0(t+is)| \le M_1(\k)\,|\sh(t+is)|^{-1/2}\ .
\label{ah.33}\end{equation} We now obtain crude bounds for $|F_m|$.
For $t \ge 0$ and $0 <|s| < \pi/2$, changing $\theta$ to
$\theta+\pi$ in (\ref{ah.8}), we get

\beq F_m(t+is) = (-1)^m \ch(t+is)^{\lambda} \int_{-\pi}^{\pi}\left (
(1-\cos(\theta))-(1-\th(t+is)) \right)^\lambda\,
e^{im\theta}\,d\theta. \label{ah.34}\end{equation} Changing to the
variable $\vhi = \theta/2$,
\begin{eqnarray}
F_m(t+is) &=& (-1)^m 2(2\ch(t+is))^{\lambda}\, A_m(z),
\label{ah.35}\\
A_m(z) &=& \int_{-\pi/2}^{\pi/2} (\sin^2(\vhi) - z^2)^{\lambda}\,
e^{2im\vhi}\,d\vhi,
\label{ah.36}\\
z^2 &=& \half(1-\th(t+is)) = \half {(1-\th t)(1-i\tg s) \over 1
+i\th t \tg s}. \label{ah.37}\end{eqnarray} We now suppose $0< \tg s
< 1/4$. It follows, after some calculations: \begin{equation} \tg s
\le \left | {\Im z^2 \over \Re z^2} \right | \le \tg(2s), \ \ \ \
\tg(s/2) \le \left | {\Im z \over \Re z} \right | \le \tg s,
\label{ah.38}\end{equation} We define $z = x-iy$ with $x > 0$. Then
\begin{equation} 0< x < |z| < 3/4,\ \ 0< y \le x\tg s < 3/16,\ \ \
{e^{-t}\over \sqrt{2}} \le |z| \le x\sqrt{1+\tg^2 s} \le
x\sqrt{17/16}. \label{ah.39}\end{equation} Recall that for $\rho >
0$, $-\pi < \theta < \pi$, $\zeta \in \bC$,
\begin{equation} |(\rho e^{i\theta})^\zeta| = \rho^{\Re
\zeta}\,e^{-\theta\,\Im \zeta} \le \rho^{\Re \zeta}\,e^{\pi|\Im
\zeta|}. \label{ah.40}\end{equation} Thus
\begin{equation} e^{-\pi|\k|} |A_m(z)| \le H(z) =
\int_{-\pi/2}^{\pi/2} |\sin^2\vhi -z^2|^{-1/2}\,d\vhi = 2\int_0^1{dt
\over \sqrt{1-t^2}\,\sqrt{|t^2-z^2|}}\ . \label{ah.41}\end{equation}
After splitting the integration interval as $[0,\ 1] = [0,\ x]\cup
[x,\ \sqrt{3}/2]\cup [\sqrt{3}/2,\ 1]$, straightforward estimates
give
\begin{equation} e^{-\pi|\k|} A_m(z) \le H(z) \le {4\pi \over 3\sqrt{3}} + 8 +
4\log(\sqrt{3}/x) \le 27+4t\ . \label{ah.41.1}\end{equation} We now
consider
\begin{equation} A_0(z) - A_m(z) = \int_{-\pi/2}^{\pi/2} (\sin^2(\vhi)
-z^2)^{\lambda} (1- e^{2im\vhi})\,d\vhi = \int_{-\pi/2}^{\pi/2}
(\sin^2(\vhi) -z^2)^{\lambda} (1- e^{2im\vhi} +2im\vhi)\,d\vhi.
\label{ah.42}\end{equation} Using \begin{equation} |1- e^{2im\vhi}
+2im\vhi| \le 2m^2\vhi^2 \le {m^2 \pi^2 \over 2}\sin^2(\vhi) \ \
\forall \vhi \in [-\pi/2,\ \pi/2] \label{ah.44}\end{equation} we get
\begin{eqnarray}
e^{-\pi|\k|} |A_0(z) - A_m(z)| &\le&
m^2 \pi^2 \int_0^{\pi/2} |\sin^2(\vhi) -z^2|^{-1/2}\,\sin^2(\vhi)\,d\vhi
\nonumber\\
&=& {m^2 \pi^2\,z^2 \over 2}H(z) +
m^2 \pi^2 \int_0^{\pi/2} |\sin^2(\vhi) -z^2|^{1/2}\,d\vhi
\nonumber\\
&\le& {m^2 \pi^2\,z^2 \over 2}H(z) + {5 \pi^3 m^2 \over 8}
\label{ah.45}\end{eqnarray} Since $|z^2| \le 17x^2/16$ and
$2x^2\log(1/x) <1/e$, there is a constant $M_2 > 0$ such that, for
all $m$, \begin{equation} |A_0(z) - A_m(z)| \le e^{\pi|\k|} M_2 m^2,
\label{ah.46}\end{equation} and hence \begin{equation} |F_0(t+is) -
(-1)^m F_m(t+is)| \le \sqrt{2}e^{2\pi|\k|} |\ch(t+is)^{-1/2-i\k}|
M_2 m^2 \label{ah.47}\end{equation} With $0 < \tg s < 1/4$, as we
have chosen, $|\ch(t+is)^{-1/2-i\k}| \le (\ch t)^{-1/2}
(17/16)^{1/4} e^{|\k|/4}$, so that \begin{equation} |F_0(t+is) -
(-1)^m F_m(t+is)| \le \sqrt{2}e^{(2\pi+1/4)|\k|}
(17/16)^{1/4}|\ch(t)|^{-1/2} M_2 m^2, \label{ah.48}\end{equation}
and, by (\ref{ah.33}), there is an $M_3(\k) > 0$ such that
\begin{equation} |F_m(t+is)| \le M_3(\k) (1+m^2) |\sh(t)|^{-1/2}.
\label{ah.49}\end{equation} Therefore, using the bound
(\ref{ah.41}), independent of $m$, for $0\le t \le t_1$, and the
bound (\ref{ah.49}) for $t_1 \le t \le T/2$, we find that there is a
constant $M_4(\k) >0$ such that \begin{equation} |T^{-1}I_m^+| \le
M_4^2(\k) (1+m^2)^2 \ \ \ \ \forall\ m \in \bZ.
\label{ah.50}\end{equation} The same holds for $|T^{-1}I_m^-|$. This
proves that Condition (2) is satisfied.

\subsection{Other dimensions}
\label{adg}

In this subsection the dimension of the de Sitter space-time $X$ is $n=d > 2$,
i.e. the ambient Minkowski space-time is $\bR^{d+1}$. The notation $n=d$
is used to stay close to \cite{Vi}, Chap. IX, p 448 ff, which
is constantly used in this section.
As in Subsect. \ref{ad2}, we wish to compute
\begin{equation}
K_\nu(\xi,\xi',g)  = \int_{X} \psi_{\lambda}^{+} (x ,\xi )\,g(x)
\,\psi_{\bar \lambda}^{-} (x ,\xi')\,dx \label{ag.1}\end{equation}
where $\xi, \xi' \in \partial V_+ \subset \bR^{d+1}$ and $\xi^0 =
{\xi'}^0 = 1$, $\lambda = -(n-1)/2 -i\nu$, $\psi_\lambda^\pm(x ,\xi
)$ are as defined in (\ref{f.13.1}). We use the following
parametrization for $x = (x^0,\ \vec{x}) \in X$, $\xi = (1,\
\vec{\xi}) \in \partial V_+$, $\xi' = (1,\ \vec{\xi'}) \in \partial
V_+$ (see \cite{Vi}, p. 448]).
\begin{equation}
\left \{
\begin{array}{l}
x^0 = \sh t,\ \ \vec{x} = -\ch t\,\vec{u}\\
u^1 = \sin\theta_{n-1}\ldots \sin\theta_2\,\sin\theta_1\\
u^2 = \sin\theta_{n-1}\ldots \sin\theta_2\,\cos\theta_1\\
u^3 = \sin\theta_{n-1}\ldots  \cos\theta_2\\
\vdots\\
u^{n-1} = \sin\theta_{n-1}\,\cos\theta_{n-2}\\
u^n = \cos\theta_{n-1}
\end{array}\right .\ \ \ \
\left \{
\begin{array}{l}
\xi^0 = 1\\
\xi^1 = 0\\
\\
\\
\vdots\\
\xi^{n-1} = 0\\
\xi^n = 1
\end{array}\right .\ \ \ \
\left \{
\begin{array}{l}
{\xi'}^0 = 1\\
{\xi'}^1 = \sin\phi_{n-1}\ldots \sin\phi_2\,\sin\phi_1\\
{\xi'}^2 = \sin\phi_{n-1}\ldots \sin\phi_2\,\cos\phi_1\\
{\xi'}^3 = \sin\phi_{n-1}\ldots  \cos\phi_2\\
\vdots\\
{\xi'}^{n-1} = \sin\phi_{n-1}\,\cos\phi_{n-2}\\
{\xi'}^n = \cos\phi_{n-1}
\end{array}\right .
\label{ag.3}\end{equation} Here $t \in \bR$, $0 \le \theta_1 <
2\pi$, $0 \le \phi_1 < 2\pi$, $0\le \theta_k <\pi$ for $k>1$, $0\le
\phi_k <\pi$ for $k>1$. With these notations
\begin{equation}
dx = \ch^{n-1}t\,dt\,d\vec{u},\ \ \ d\vec{u} =
\sin^{n-2}\theta_{n-1}\,d\theta_{n-1}\ldots
\sin\theta_2\,d\theta_2\,d\theta_1\ . \label{ag.4}\end{equation} We
also use the normalized measure $d\sigma(\vec{u})$ on $S^{n-1}$,
\begin{equation}
d\sigma(\vec{u}) = \Omega_n^{-1}\,d\vec{u},\ \ \ \Omega_n =
\int_{S^{n-1}} d\vec{u} = {2 \pi^{n/2} \over \Gamma(n/2)}\ .
\label{ag.5}\end{equation} We restrict $g$ to be of the form $g(x) =
g_T(x) = g_0(t) = \theta(T/2-t)\theta(t+T/2)$, $T>0$. The integral
(\ref{ag.1}) takes the form
\begin{equation}
K_\nu(\xi,\xi',g)  = \int_{\bR} g_0(t)\,(\ch t)^{n-1}\,dt \int F(t,\
\vec{u})\,\ovl{G(t,\ \vec{u})}\,d\vec{u}\ ,
\label{ag.6}\end{equation}

\beq F(t,\ \vec{u}) = (\sh(t+i0) +
\ch(t+i0)\,\cos(\theta_{n-1}))^\lambda\ . \label{ag.7}\end{equation}
For $G$, we have $G(t,\ \vec{u}) = (x_+\cdot \xi')^\lambda$. Note
that $\xi' = R\xi$, where $R$ is the rotation in $\bR^n$
\begin{equation}
R = e^{\phi_1\,M_{21}}\ldots e^{\phi_{n-1}\,M_{n n-1}},\ \ \ M_{jk}
= e_j\wedge e_k\ . \label{ag.8}\end{equation} For example
\begin{equation}
e^{\phi_{n-1}\,M_{n n-1}} = \left (
\begin{array}{ccccc}
1 & \ldots & 0 & 0 & 0\\
\vdots &&&& \vdots\\
0 & \ldots & 1 & 0 & 0\\
0 & \ldots & 0 &\cos \phi_{n-1} & \sin \phi_{n-1}\\
0 & \ldots & 0 & -\sin \phi_{n-1} & \cos \phi_{n-1}
\end{array}
\right )\ . \label{ag.8.1}\end{equation} Therefore
\begin{equation}
G(t,\ \vec{u}) = F(t,\ R^{-1}\vec{u}). \label{ag.9}\end{equation} As
in the case $d=2$, we reexpress the integral over $\vec{u}$ in
(\ref{ag.6}) using harmonic analysis on the sphere.

Harmonic analysis on $S^{n-1}$ uses an orthonormal basis
$\{\Xi^\ell_K\}$ of functions on the sphere ($\ell = 0,\ 1,\ 2,\ \ldots$,
$K$ is a multiindex).
This is fully described in \cite{Vi}, Chap IX]):
\begin{equation}
\int_{S^{n-1}} \Xi^\ell_K(\vec{u})\,
\ovl{\Xi^{\ell'}_{K'}(\vec{u})}\,d\sigma(\vec{u}) = \delta_{\ell
\ell'}\delta_{K K'}\ . \label{ag.9.1}\end{equation} For fixed $\ell$
the functions $\{\Xi^\ell_K\}$ generate a finite-dimensional
subspace $\HH^{n\ell}$ of $L^2(S^{n-1})$ in which the regular
representation of $SO(n)$ reduces to an irreducible unitary
representation, characterized by its matrix elements in the basis
$\{\Xi^\ell_K\}$: for any $g \in SO(n)$,
\begin{equation}
\Xi^\ell_K(g^{-1}\vec{u}) = \sum_M
t^\ell_{MK}(g)\,\Xi^\ell_M(\vec{u})\ . \label{ag.9.2}\end{equation}
The functions $\{\Xi_K^\ell\}$ and $\{t_{KM}^\ell\}$ are analytic on
$S^{n-1}$ and $SO(n)$ respectively. Given two arbitrary $L^2$
functions $h_1$, $h_2$ on $S^{n-1}$ we have (for $j = 1,\ 2$)
\begin{eqnarray}
h_j(\vec{u}) &=& \sum_{\ell,\ K} {h_j}_K^\ell\,\Xi^\ell_K(\vec{u}),
\label{ag.10}\\
{h_j}_K^\ell &=&
\int_{S^{n-1}} h_j(\vec{u})\,\ovl{\Xi^\ell_K(\vec{u})}\,d\sigma(\vec{u}),
\label{ag.11}\\
\int_{S^{n-1}} h_1(\vec{u})\,\ovl{h_2(\vec{u})}\,d\sigma(\vec{u})
&=& \sum_{\ell,\ K} {h_1}_K^\ell\,\ovl{{h_2}_K^\ell}\ .
\label{ag.12}\end{eqnarray} These formulae imply that
\begin{equation} \sum_{\ell,\ K}
\ovl{\Xi_K^\ell(\vec{u})}\,\Xi_K^\ell(\vec{v}) =
\Omega_n\delta_{S^{n-1}}(\vec{u},\ \vec{v})
\label{ag.12.1}\end{equation} where $\delta_{S^{n-1}}(\vec{u},\
\vec{v})$ denotes the distribution (actually measure) on
$S^{n-1}\times S^{n-1}$ defined by
\begin{equation} \int_{S^{n-1}\times S^{n-1}}
\delta_{S^{n-1}}(\vec{u},\ \vec{v})\, \vhi(\vec{u},\
\vec{v})\,d\vec{u}\,d\vec{v} = \int_{S^{n-1}} \vhi(\vec{u},\
\vec{u})\,d\vec{u}\ . \label{ag.12.2}\end{equation} Actually, as is
the case for all invariant distributions on $S^{n-1}\times S^{n-1}$,
smearing $\delta_{S^{n-1}}(\vec{u},\ \vec{v})$ only in $\vec{v}$
with a $\CC^\infty$ function produces a $\CC^\infty$ function of
$\vec{u}$:
\begin{equation} \int_{S^{n-1}} \delta_{S^{n-1}}(\vec{u},\
\vec{v})\,\psi(\vec{v})\, d\vec{v} = \psi(\vec{u})\ .
\label{ag.12.3}\end{equation} Choosing in particular $\vec{u} =
\vec{e_n}$, we can use the formula (\cite{Vi}, IX 4.1 (1-4) and text
there)
\begin{equation} \Xi_K^\ell(\vec{e_n}) = \delta_{K0}
\sqrt{\Gamma(\ell+n-2)(2\ell+n-2)\over \ell!\,\Gamma(n-1)}.
\label{ag.12.4}\end{equation} Inserting this in (\ref{ag.12.1})
gives
\begin{equation} \delta_{S^{n-1}}(\vec{e_n},\ \vec{v}) =
\Omega_n^{-1}\sum_{\ell} \sqrt{\Gamma(\ell+n-2)(2\ell+n-2)\over
\ell!\,\Gamma(n-1)}\, \Xi_0^\ell(\vec{v}).
\label{ag.12.5}\end{equation} Taking $\vec{v} = \vec{\xi'}$ with
$\vec{\xi'}$ given by (\ref{ag.3}) and using \cite{Vi} IX 4.1 (3-4),
\begin{equation} \Xi_M^\ell(g\vec{e_n}) =
\sqrt{\Gamma(\ell+n-2)(2\ell+n-2)\over \ell!\,\Gamma(n-1)}\,
t_{M0}^\ell(g)
\end{equation}
we get
\begin{equation}
\delta_{S^{n-1}}(\vec{e_n},\ \vec{\xi'}) = \Omega_n^{-1}\sum_{\ell}
{\Gamma(\ell+n-2)(2\ell+n-2)\over
\ell!\,\Gamma(n-1)}\,t_{00}^\ell(R), \label{ag.12.6}\end{equation}
with $R$ given by (\ref{ag.8}). Harmonic analysis extends to
distributions on the sphere, as it does on $S^1$. We apply
(\ref{ag.10}-\ref{ag.12}) to the case $h_1(\vec{u}) = F(t,\
\vec{u})$, $h_2(\vec{u}) = G(t,\ \vec{u})$. Because $F(t,\ \vec{u})$
depends only on $\cos\theta_{n-1}$,
\begin{equation}
f^\ell_K(t) = \int_{S^{n-1}} F(t,\
\vec{u})\,\ovl{\Xi^\ell_K(\vec{u})}\,d\sigma(\vec{u}) =\delta_{K0}
f^\ell_0(t)\ . \label{ag.14}\end{equation} Note that $t$ can be
complexified in (\ref{ag.14}), i.e. $t$ can be replaced by $t+is$
with $0<|s|<\pi/2$. In the sequel we omit the $t$-dependence of
$f^\ell_K(t)$, writing simply $f^\ell_K$ unless the $t$-dependence
becomes significant. We have
\begin{equation}
\Xi^\ell_0(\vec{u}) = A^\ell_0 C_\ell^{n-2\over
2}(\cos\theta_{n-1}), \ \ \ A^\ell_0 =
\sqrt{\ell!\Gamma(n-2)(2\ell+n-2) \over \Gamma(\ell+n-2)(n-2)}\ .
\label{ag.15}\end{equation}
\begin{equation}
G(t,\ \vec{u}) = \sum_\ell f^\ell_0\,\Xi^\ell_0(R^{-1}\vec{u}) =
\sum_{\ell,\ K} (f^\ell_0\,t^\ell_{K0}(R))\,\Xi^\ell_K(\vec{u})\ .
\label{ag.16}\end{equation} Therefore
\begin{eqnarray}
\int_{S^{n-1}} F(t,\ \vec{u})\,\ovl{G(t,\ \vec{u})}\,d\vec{u} &=&
\Omega_n\,\sum_\ell t^\ell_{00}(R)\,|f^\ell_0|^2\,,
\label{ag.17}\\
K_\nu(\xi,\xi',\ g_T) &=&  \Omega_n\,\sum_\ell t^\ell_{00}(R)\,
\int_{-T/2}^{T/2} |f^\ell_0(t)|^2\,(\ch t)^{n-1}\,dt
\label{ag.17a}\end{eqnarray} Also \begin{equation} t^\ell_{00}(R) =
{\ell! \Gamma(n-2) \over \Gamma(\ell +n -2)} C_\ell^{n-2\over
2}(\cos\phi_{n-1})\ , \label{ag.18}\end{equation} For (\ref{ag.15})
see \cite{Vi} IX 3.6 (6,7) p. 480. For (\ref{ag.18}) see \cite{Vi},
IX 4.2 (8) p. 484. We thus have
\begin{eqnarray}
f^\ell_0 &=& \Omega_n^{-1} A_0^\ell
\int [\sh(t+i0) + \ch(t+i0)\,\cos\theta_{n-1}]^\lambda\,
C_\ell^{n-2\over 2}(\cos\theta_{n-1})
\sin^{n-2}\theta_{n-1}\,d\theta_{n-1}\ldots
\sin\theta_2\,d\theta_2\,d\theta_1\nonumber\\
&=& \Omega_n^{-1} \Omega_{n-1} A_0^\ell \int_0^\pi [\sh(t+i0) +
\ch(t+i0)\,\cos\theta]^\lambda\, \sin^{n-2}\theta\, C_\ell^{n-2\over
2}(\cos\theta)\,d\theta \label{ag.20}\end{eqnarray} In these
formulae $C_\ell^\mu$ is a Gegenbauer polynomial: see \cite{bateman1} p.
175 for the definition. The formulae \cite{bateman1} p. 176 (9), and
\cite{Vi}, IX 3.1 (3), giving the explicit coefficients of
$C_\ell^\mu$ coincide, so we are dealing with the same objects.

\subsection{The case $d = n = 4$}
\label{n4}

We now restrict our attention to the case $d=4$,
keeping the notations of the preceding subsection.
In this case $\lambda = -3/2-i\k$, $\Omega_4 = 2\pi^2$,
$A_0^\ell = 1$. We exclude the case $\k=0$.
Since $(n-2)/2 = 1$, the formula (\ref{ag.20}) gives:
\begin{equation}
f_0^\ell = {2\over\pi} \int_0^\pi [\sh(t+i0) +
\ch(t+i0)\cos\theta]^\lambda\,
C_\ell^1(\cos\theta)\,\sin^2\theta\,d\theta.
\label{n4.1}\end{equation} We have (\cite{bateman1}, 3.15.1 (15) p. 177)
\begin{equation}
C_\ell^1(\cos\theta) = {\sin(\ell+1)\theta \over \sin\theta}\ .
\label{n4.5}\end{equation} Therefore, for sufficiently small $s>0$,
\begin{eqnarray}
f_0^\ell(t+is) &=& {2\over\pi}
\int_0^\pi [\sh(t+is) + \ch(t+is)\cos\theta]^\lambda\,
\sin(\ell+1)\theta\,\sin\theta\,d\theta
\label{n4.7}\\
&=& {1\over\pi}\ch(t+is)^\lambda
\int_{-\pi}^\pi [\th(t+is) + \cos\theta]^\lambda\,
\sin(\ell+1)\theta\,\sin\theta\,d\theta
\label{n4.8}\\
&=& {(\ell+1)\over\pi(\lambda+1)}\ch(t+is)^\lambda
\int_{-\pi}^\pi [\th(t+is) + \cos\theta]^{\lambda+1}\,
\cos(\ell+1)\theta\,d\theta
\label{n4.9}\\
&=& {(\ell+1)\over\pi(\lambda+1)}\ch(t+is)^{-1}
\int_{-\pi}^\pi [\sh(t+is) + \ch(t+is)\cos\theta]^{\lambda+1}\,
\cos(\ell+1)\theta\,d\theta\ .
\label{n4.10}
\end{eqnarray}
Recall that
$\lambda +1 = -1/2-i\k$.
Therefore, comparing (\ref{n4.10}) with (\ref{ah.8}), we find,
for $0<|s|<\pi/2$,
\begin{eqnarray}
f_0^\ell(t+is) &=& {(\ell+1)\over 2\pi(-\half-i\k)\ch(t+is)} \left(
F_{\ell+1}(t+is) + F_{-(\ell+1)}(t+is) \right)\cr &=& {(\ell+1)\over
\pi(-\half-i\k)\ch(t+is)}\,F_{\ell+1}(t+is)\ ,
\label{n4.11}\end{eqnarray} using (\ref{ah.8.1}). Hence
\begin{eqnarray}
K_\k(\xi,\xi',g_T)  &=& \Omega_4\sum_\ell t_{00}^\ell(R)\,
{(\ell+1)^2 \over \pi^2(\k^2 +1/4)}\,
\int_{-T/2}^{T/2} F_{\ell+1}(t+i0)\,\ovl{F_{\ell+1}(t+i0)}\,\ch t\,dt
\label{n4.14}\\
&=& \sum_\ell t_{00}^\ell(R)\, {4\pi(\ell+1)^2 \over (\k^2
+1/4)}\,(I_{\ell+1}^+ + I_{\ell+1}^-) \label{n4.15}\end{eqnarray}
with the notations of (\ref{ah.16}). Therefore, by (\ref{ah.29}) and
the proof of Condition (2) for $d=2$ (subsect. \ref{cond2}),
\begin{equation} \lim_{T\rightarrow +\infty} {1\over T}
K_\k(\xi,\xi',g_T) = {8\pi e^{\pi\k}\,\ch(\pi\k)^2 \over (\k^2
+1/4)\k\,\sh(\pi\k)}\, \sum_\ell t_{00}^\ell(R)\,(\ell+1)^2\ .
\label{n4.16}\end{equation} In the case $n=4$, (\ref{ag.12.6})
becomes
\begin{equation} \sum_\ell (l+1)^2\,t_{00}^\ell(R) =
\Omega_4\delta_{S^3}(\vec{e_4},\ \vec{\xi'}) = 2\pi^2
\delta_{S^3}(\vec{\xi},\ \vec{\xi'})\ . \label{n4.17}\end{equation}
Thus
\begin{equation} \lim_{T\rightarrow +\infty} {1\over T}
K_\k(\xi,\xi',g_T) = {16\pi^3 e^{\pi\k}\,\ch(\pi\k)^2 \over (\k^2
+1/4)\k\,\sh(\pi\k)}\,\delta_{S^3}(\vec{\xi},\ \vec{\xi'})\ .
\label{n4.19}\end{equation} It follows (see (\ref{ah.1}) that
\begin{equation} \lim_{T\rightarrow +\infty} T^{-1}L_1(f_0,\ g_T) =
{\coupl^2 \pi \coth(\pi\k)^2 \over |\k|}\ .
\label{n4.20}\end{equation} This is the same as in the case $d=2$.

\section{Appendix. Adiabatic limit (dS): parabolic slices}

\label{adpar}
We again take $R=1$.
We again start from the formulae (\ref{ah.1}-\ref{ah.3}) of
Appendix \ref{adhor}, but we only require
$\k \in \bR\setminus \{0\}$.
The function $g$ will be chosen as announced in subsect \ref{dSadlim}.
The map $(t,\ \y) \mapsto x(t,\ \y)$ defined in (\ref{s.4})
is a diffeomorphism of $\bR^d$ onto the ``upper half''
$X_d^{\rm up} = \{x\in X_d\ :\ x^0+x^d > 0\}$, and $(t,\ \y) \mapsto -x(t,\ \y)$
is a diffeomorphism of $\bR^d$ onto the ``lower half''
$X_d^{\rm down} = -X_d^{\rm up}$. The cycle $\gamma$ appearing in
(\ref{ah.1}) will be chosen as
$V_0 = C_+\cap \{\xi\in M_{d+1}\ :\  \xi^0+\xi^d = 1\}$. It can be parametrized
by the diffeomorphism $\eta \mapsto \xi(\eta)$ of $\bR^{d-1}$ onto $V_0$ :
\begin{equation}
\xi(\eta) = \left\{\begin{array}{lcl}
\xi^0 &=& \half (1+ \eta^2),\\
\xi^{j} &=&  \eta_{j},\ \ (1\le j \le d-1)\,,\\
\xi^d &=& \half (1- \eta^2),
\end{array}\right . \ \ \ \eta^2 = \sum_{j=1}^{d-1} \eta_j^2\ .
\label{ap.1}\end{equation}
Thus $V_0$ is a Euclidean space with
$(d\xi\cdot d\xi) = -d\eta^2$ on $V_0$. The stability group of the
vector $e_0-e_d$ in $G_0$ leaves $V_0$ invariant and acts as the
group of Euclidean displacements there. As noted in Remarks \ref{homg}
and \ref{psipm}, the $G_0$ invariance and
homogeneity of $\psi_\lambda^\pm(x,\ \xi)$ imply that it can
be regarded as a distribution in $\xi$ on $V_0$, $\CC^\infty$ in $x$
on $X_d$.
For a real $g\in \SS(X_d)$, if we denote
$\check g(x) = g(-x)$, we find
\begin{equation}
K_\k(\xi,\xi',\check g) = e^{2\pi\k}\ovl{K_{-\k}(\xi,\xi',g)}\ .
\label{ap.2}\end{equation}
It will turn out that $g$ can be chosen invariant under
the stability group of $e_0-e_d$. Then $K_\k$ is an invariant
distribution on $V_0\times V_0$.
For our purposes it will suffice (and be possible) to
study the limit of $T^{-1}K_\k(\xi,\xi',g_T)$ with
\begin{equation}
g_T(x) = \theta(t+T/2)\theta(T/2-t),\ \ \ t = \log(x^0+x^d)\ ,
\label{ap.3}\end{equation}
and to add in the end the limit of
$T^{-1}K_\k(\xi,\xi',\check g_T)$ obtained from (\ref{ap.2}).

With $x$ parametrized as in (\ref{s.4}) and $\xi$ as in (\ref{ap.1}),
we have
\begin{equation}
x(t,\ \y)\cdot \xi = \half [e^t(\y-\eta)^2-e^{-t}] ={(\y-\eta)^2
\over 2s} - {s\over 2},\ \ \ s = e^{-t}\ .
\label{ap.4}\end{equation} For $\kr \in \bR^{d-1}$, we find
\begin{eqnarray}
\wt{\psi_{-{d-1\over 2}+i\nu}^\pm}(\kr,\ s,\ \eta) &\bydef&
\int_{\bR^{d-1}} \psi_{-{d-1\over 2}+i\nu}^\pm(x(t,\ \y),\ \xi)\,
e^{i\kr\y}\,d\y\cr &=& 2^{{d-1\over
2}-i\nu}e^{i\kr\eta}\int_{\bR^{d-1}} \left [ {\y^2\over s\mp
i\epsilon} -(s\mp i\epsilon)\right ]^{-{d-1\over
2}+i\nu}\,e^{i\kr\y}\,d\y\cr &=& 2^{{d-1\over
2}-i\nu}e^{i\kr\eta}\int_0^\infty \left [ {a^2\over s\mp i\epsilon}
-(s\mp i\epsilon)\right ]^{-{d-1\over 2}+i\nu}\,da^2\,
\int_{\bR^{d-1}} \delta(\y^2-a^2)\,e^{i\kr\y}\,d\y\cr &=&
2^{d-1-i\nu}\pi^{d-1\over 2}k^{3-d\over 2}e^{i\kr\eta}\,
\int_0^\infty \left [ {y^2\over s\mp i\epsilon} -(s\mp
i\epsilon)\right ]^{-{d-1\over 2}+i\nu}\,y^{d-1\over 2} J_{d-3\over
2}(ky)\,dy\ ,\hbox to 1cm{\hfill} \label{ap.5}\end{eqnarray} where
$k = |\kr|$ and $s=e^{-t}$.
We use the following formula (\cite{bateman2}
 (51) p. 95
with some notational changes)
\begin{equation}
\int_0^\infty (y^2 +z^2)^{-\lambda+i\alpha} y^\lambda
J_{\lambda-1}(ky)\,dy = \left({k\over 2}\right)^{\lambda-i\alpha-1}
{z^{i\alpha}K_{-i\alpha}(kz)\over \Gamma(\lambda -i\alpha)}\ .
\label{ap.6}\end{equation}
This is valid provided $k>0$, $\Re z >
0$, $\Re \lambda >0$, and $\Re (\lambda-2i\alpha+1/2) >0$. Note that
none of these parameters except $k$ needs to be real. In our
application, $\lambda = (d-1)/2$ and $\alpha = \nu \in \bR\setminus
\{0\}$. We take $z= -is+\veps$, $s>0$, $\veps >0$ arbitrarily small,
$\alpha = \nu$. Since $(y^2+z^2)$ then has a small negative
imaginary part, this will correspond to the case of
$\wt{\psi_{-\lambda+i\nu}^-}$ in (\ref{ap.5}).
In (\ref{ap.6}) $K_{-i\alpha}$ denotes the Macdonald function.
This will introduce no lasting ambiguity since
we will use the identities
(\cite{bateman2}, (5), (6) p.~4, (15) p.~5)
\begin{equation}
K_{-i\nu}(-iks) = {i\pi\over 2}e^{\pi\nu/2} H_{-i\nu}^{(1)}(ks) =
{i\pi e^{\pi\nu/2}\over 2\sh(\pi\nu)} \left ( J_{i\nu}(ks) -
e^{-\pi\nu}J_{-i\nu}(ks)\right ).
\label{ap.7}\end{equation}
This yields
\begin{equation}
\wt{\psi_{-\lambda+i\nu}^-}(\kr,\ s,\ \eta) = e^{i\kr\eta}
{i2^\lambda \pi^{\lambda+1}k^{-i\nu}s^\lambda\over
\Gamma(\lambda-i\nu)\sh(\pi\nu)} \left ( e^{\pi\nu}J_{i\nu}(ks) -
J_{-i\nu}(ks) \right ),\ \ \ \lambda = {d-1\over 2}\  .
\label{ap.8}\end{equation}
$\wt{\psi_{-\lambda-i\nu}^+}(\kr,\ s,\
\eta)$ can be obtained from this since, for real $\nu$, it is the
complex conjugate of $\wt{\psi_{-\lambda+i\nu}^-}(-\kr,\ s,\ \eta)$:
\begin{equation}
\wt{\psi_{-\lambda-i\nu}^-}(\kr,\ s,\ \eta) = e^{i\kr\eta}
{(-i)2^\lambda \pi^{\lambda+1}k^{i\nu}s^\lambda\over
\Gamma(\lambda+i\nu)\sh(\pi\nu)} \left ( e^{\pi\nu}J_{-i\nu}(ks) -
J_{i\nu}(ks) \right ),\ \ \ \lambda = {d-1\over 2}\  .
\label{ap.8.1}\end{equation}
\begin{remark}\rm
\label{psirem}
By the preceding remarks, if $\xi \in V_0$
is expressed in terms of $\eta \in \bR^{d-1}$ as in (\ref{ap.1}),
$\psi_{\alpha}^{\pm} (x,\xi)$ is a tempered distribution in $\eta$,
a $\CC^\infty$ function of $x$, and an entire function in $\alpha$.
If $x$ is expressed as in (\ref{s.4}),
its Fourier transform with respect to the variable $\y$ is also a tempered
distribution in the variable $\kr$ conjugated to $\y$ and in
$\eta$, $\CC^\infty$ in $s$ and holomorphic in $\alpha$, and,
in this sense, the formulae (\ref{ap.8}) and (\ref{ap.8.1})
can be continued to all $\nu$. If $\nu$ is taken real in these formulae,
their rhs becomes locally bounded in $\kr$, in particular locally $L^2$.
\end{remark}
Supposing $g(x) = G((x^0+x^d)^{-1})$ (for example if
$g(x) = g_T(x) = G_T(s) = \theta(s-e^{-T/2})\theta(e^{T/2}-s)$),
Plancherel's formula gives
\begin{equation}
K_\k(\xi,\xi',g) = (2\pi)^{1-d}\int_{s>0,\ \kr \in \bR^{d-1}} s^{-d}
G(s)\, \wt{\psi_{-{d-1\over 2} -i\k}^{+}}(\kr,\ s,\ \eta)
\wt{\psi_{-{d-1\over 2} +i\k}^{-}}(-\kr,\ s,\ \eta')\,ds\,d\kr\ .
\label{ap.9}\end{equation} Inserting (\ref{ap.8}), we find
\begin{equation}
K_\k(\xi,\xi',g) = (2\pi)^{1-d}\int_{\bR^{d-1}} e^{i\kr\cdot
(\eta-\eta')} \wt{K_\k}(\kr,\ g)\,d\kr, \label{ap.10}\end{equation}
\begin{equation}
\wt{K_\k}(\kr,\ g) = \int_0^\infty s^{-1}G(s) \left
[A\,J_{i\k}(ks)J_{-i\k}(ks) + BJ_{i\k}^2(ks) + CJ_{-i\k}^2(ks)
\right ]\,ds\ , \label{ap.11}\end{equation} where
\begin{equation}
A= {2^{d}\pi^{d+1}e^{\pi\k}\ch(\pi\k)\over \Gamma\left({d-1\over
2}+i\k\right)\Gamma\left({d-1\over 2}-i\k\right) \sh^2(\pi\k)},
\label{ap.11.1}\end{equation}
\begin{equation}
B = C = {-2^{d-1}\pi^{d+1}e^{\pi\k}\over \Gamma\left({d-1\over
2}+i\k\right)\Gamma\left({d-1\over 2}-i\k\right) \sh^2(\pi\k)}\ .
\label{ap.11.2}\end{equation}

Going back to eq. (\ref{ap.11}), we divide the integration
range into the intervals $[0,\ 1]$ and $[1,\ \infty]$. After dividing by
$T$, the contribution of the second interval is bounded by
 \begin{equation}
\hbox{Const.} {1\over T}\int_1^\infty k^{-1}s^{-2}\,ds = {\rm
Const.} {1\over kT}. \label{ap2.1}\end{equation} This is because
$|J_\alpha(x)| < \hbox{Const.}x^{-1/2}$ as $x\rightarrow +\infty$
(see \cite{bateman2}, p. 85). Hence the contribution of the second
interval tends to 0 as $T$ tends to $+\infty$. The function
$J_\alpha$ can be written as
\begin{equation}
J_\alpha(z) = (z/2)^\alpha \sum_{m=0}^\infty {(-1)^m(z/2)^{2m}\over
m!\Gamma(m+\alpha+1)} = (z/2)^\alpha \left ({1\over
\Gamma(1+\alpha)}+O(z^2) \right ). \label{ap2.2}\end{equation} Thus
as $T$ tends to $+\infty$,
\begin{eqnarray}
&&{1\over T}\wt{K_\k}(\kr,\ g_T) \sim {1\over T}\int_{e^-{T/2}}^1
s^{-1} \left [ A\,J_{i\k}(ks)J_{-i\k}(ks) + BJ_{i\k}^2(ks) +
CJ_{-i\k}^2(ks) \right ]\,ds\ \sim\cr &&{1\over T}\int_{e^-{T/2}}^1
s^{-1} \left [ {A\over \Gamma(1+i\k)\Gamma(1-i\k)}
+{B(ks/2)^{2i\k}\over \Gamma(1+i\k)^2} + {C (ks/2)^{-2i\k}\over
\Gamma(1-i\k)^2} + \hbox{Const.} k^2s^2 \right ]\,ds\cr && = {A\over
2\Gamma(1+i\k)\Gamma(1-i\k)} + {Bk^{2i\k}(1-e^{-iT\k})\over
2iT\k\Gamma(1+i\k)^2} + {Ck^{-2i\k}(1-e^{iT\k})\over
-2iT\k\Gamma(1-i\k)^2} + {\hbox{Const.}k^2(1-e^{-T})\over 2T}\ .
\label{ap2.3}\end{eqnarray} Hence \begin{equation} \lim_{T
\rightarrow +\infty} {1\over T}\wt{K_\k}(\kr,\ g_T) = {A\over
2\Gamma(1+i\k)\Gamma(1-i\k)} = {A\sh(\pi\k)\over 2\pi\k}\ .
\label{ap2.4}\end{equation} This gives \begin{equation} \lim_{T
\rightarrow +\infty} {1\over T}K_\k(\xi,\xi',g_T) =
{A\sh(\pi\k)\over 2\pi\k}\,\delta(\eta-\eta'),
\label{ap2.5}\end{equation} and (see (\ref{ah.1})) \begin{equation}
\lim_{T \rightarrow +\infty} {1\over T}L_1(f_0,\ g_T) = {\coupl^2
C_0(\k) c_{d,\k} A\sh(\pi\k)\over 2\pi\k} = {\coupl^2
\pi\coth^2(\pi\k) \over 2|\k|}. \label{ap2.6}\end{equation} This is
half of the result in (\ref{ah.31}) or (\ref{n4.20}), but it is
doubled by the addition of the contribution of $\check g_T$.

\section{Proof of the projector identity}
\label{projector}
In this appendix, we give a proof of the formula (\ref{f.18})
in the de Sitter case,
with masses $m$ and $m'$ in the principal series, i.e.
$m^2 = \mu^2+(d-1)^2/4$, $m'^2 = \nu^2 +(d-1)^2/4$, with real $\mu\not=0$ and
$\nu \not=0$. We set $R$ equal to 1.
The meaning of (\ref{f.18}) is
\beq
\lim_{g\in \SS(X_d),\ g \rightarrow 1}
\int_\XX \w_{m}(z,\ x)\,\w_{m'}(x,\ y)\,g(x)\,dx =
C_1(m,\ d)\delta(m^2-m^{\prime 2}) \w_{m}(z,\ y).
\label{i.0}\endq
For $g\in \SS(X_d)$ the integral in this formula is well defined
(see Remark \ref{psipm}).
The same method as in Appendix \ref{adpar} will be used.
Using Eq. (\ref{f.16.2}) reduces
the problem to the study, as $g$ tends to 1, of
\beq
K_{\mu,\nu}(\xi,\xi',g) = \int_{X_d} \,
\psi_{-{d-1\over 2} -i\mu}^{+} (x,\xi)\,
\psi_{-{d-1\over 2} +i\nu}^{-} (x ,\xi')\,g(x)\,dx\ .
\label{i.1}\endq
Recalling Remarks \ref{homg}, \ref{psipm} and \ref{psirem},
and parametrizing $\xi$ and $\xi'$ in terms of $\eta$ and $\eta'$
as in (\ref{ap.1}), we see that,
for a general smooth fast decreasing $g$, this
is well defined as a distribution in $\eta$ and $\eta'$,
and an entire function in $\mu$ and $\nu$,
and, denoting $\check g(x) = g(-x)$, it satisfies
\begin{equation}
K_{\mu,\nu}(\xi,\xi',\check g) = e^{\pi(\mu+\nu)}
\ovl{K_{-\bar\mu,-\bar\nu}(\xi,\xi',\ovl{g})}\ .
\label{i.1.1}\end{equation}
(It is sufficient to verify this formula for real
$\mu$ and $\nu$).
We will use the same coordinates (\ref{s.4}) and
many of the formulae of Appendix \ref{adpar}.
We wish to take $g$ as $g_u(x)=\theta(x^0+x^d)$, or $g_d= \check g_u$.
Thus $g_u$ ($u$ stands for ``upper'') is the indicator
function of the domain covered by the coordinates (\ref{s.4}),
We denote $K_{\mu,\nu}^u(\xi,\xi') = K_{\mu,\nu}(\xi,\xi',g_u)$ and
$K_{\mu,\nu}^d(\xi,\xi') = K_{\mu,\nu}(\xi,\xi',\check g_u)$.
To make the integral converge, we first replace
$g_u$ by a better behaved $g_u^\veps$ of the form
$g_u^\veps(x(t,\ \y)) = G_\veps(e^{-t})g_u(x)$ which will tend to $g_u(x)$
as $\veps \rightarrow 0$.
We thus consider
\beqa
&&K_{\mu,\nu}^{u,\veps}(\xi,\xi') =
K_{\mu,\nu} (\xi,\xi', g_u^\veps)=
\int_{X_d^u} \,
\psi_{-{d-1\over 2} -i(\mu)}^{+} (x,\xi)\,
\psi_{-{d-1\over 2} +i\nu}^{-} (x ,\xi')\,g_u^\veps(x)\,dx\ ,
\label{i.1bis}\\
&&K_{\mu,\nu}^{d,\veps}(\xi,\xi') =
K_{\mu,\nu} (\xi,\xi', \check g_u^\veps)= e^{\pi(\mu+\nu)}
\ovl{K_{-\bar\mu,-\bar\nu}^{u,\veps}(\xi,\xi')}\ .
\label{i.1.1bis}\endqa
We now take $\mu$ and $\nu$ real and furthermore require $\mu\nu>0$.
Using the coordinates
(\ref{s.4}) and parametrizing $\xi$ and $\xi'$ as in Appendix \ref{adpar}
(see (\ref{ap.1})), we may use the Plancherel formula as was done there.
We obtain
\beqa
&&K_{\mu,\nu}^{u,\veps}(\xi,\xi') = (2\pi)^{1-d}
\int_{\bR^{d-1}} e^{i\kr\cdot(\eta-\eta')}\,
\wt{K_{\mu,\nu}^{u,\veps}}(\kr)\,d\kr\ ,
\label{i.2}\\
&&K_{\mu,\nu}^{d,\veps}(\xi,\xi') = (2\pi)^{1-d}
\int_{\bR^{d-1}} e^{i\kr\cdot(\eta-\eta')}\,
\wt{K_{\mu,\nu}^{d,\veps}}(\kr)\,d\kr\ ,\ \ \
\wt{K_{\mu,\nu}^{d,\veps}}(\kr) =
e^{\pi(\mu+\nu)}\ovl{\wt{K_{-\mu,-\nu}^{u,\veps}}(-\kr)}\ ,
\label{i.2.1}\endqa
\begin{eqnarray}
&& \wt{K_{\mu,\nu}^{u,\veps}}(\kr) =  {2^{d-1}\pi^{d+1} k^{i(\mu-\nu)}\over
\sh (\pi \mu)\,\sh (\pi \nu)\,
\Gamma\left(\frac{d-1}{2} - i\,\mu \right)
\Gamma\left(\frac{d-1}{2} + i\,\nu
\right)}
\times \cr
&& \times \int_0^\infty{ds\over s}\, G_\veps(s)
\left[e^{\pi\mu} {J}_{-i\mu} (sk) - J_{i\mu}(sk) \right]
\left[e^{\pi\nu} {J}_{i\nu} (sk) - J_{-i \nu}(sk) \right],
\label{i.3}\end{eqnarray}
where $k=|\kr|$.
We can use the following formula (\cite{bateman2}, 7.7.4 (30), p. 51):
\begin{align}
&\int_0^\infty J_\alpha(as)J_\beta(as)\,s^{-\rho}\,ds =\cr
&{(a/2)^{\rho-1} \Gamma(\rho)\Gamma((\alpha+\beta+1-\rho)/2) \over
2\Gamma((1+\alpha+\beta+\rho)/2) \Gamma((1-\alpha+\beta+\rho)/2)
\Gamma((1+\alpha-\beta+\rho)/2)}\ ,\cr
&
\Re(\alpha+\beta+1) > \Re \rho > 0,\ \ \ \ a>0\ .
\label{i.4}\end{align}
Choosing $G_\veps(s) = s^\veps$ with
$0<\veps <1$,  and using (\ref{i.2.1}) to obtain $\wt{K_{\mu,\nu}^{d,\veps}}(\kr)$
from $\wt{K_{\mu,\nu}^{u,\veps}}(\kr)$, we obtain
\begin{align}
& \wt{K_{\mu,\nu}^{u,\veps}}(\kr)  + \wt{K_{\mu,\nu}^{d,\veps}}(\kr) =
{2^{d-2}\pi^{d+1} k^{i(\mu-\nu)}(k/2)^{-\veps}\Gamma(1-\veps)\over
\sh (\pi \mu)\,\sh (\pi \nu)\,
\Gamma\left(\frac{d-1}{2} - i\,\mu \right)
\Gamma\left(\frac{d-1}{2} + i\,\nu
\right)}
\times \cr
& \left [
{(e^{\pi(\mu+\nu)}+1) \Gamma((-i\mu+i\nu+\veps)/2) \over
\Gamma((2-i\mu+i\nu-\veps)/2)\Gamma((2+i\mu+i\nu-\veps)/2)
\Gamma((2-i\mu-i\nu-\veps)/2)}\right .\cr
&- {(e^{\pi\mu}+e^{\pi\nu})\Gamma((-i\mu-i\nu+\veps)/2) \over
\Gamma((2-i\mu-i\nu-\veps)/2)\Gamma((2+i\mu-i\nu-\veps)/2)
\Gamma((2-i\mu+i\nu-\veps)/2)} \cr
&- {(e^{\pi\mu}+e^{\pi\nu})\Gamma((i\mu+i\nu+\veps)/2) \over
\Gamma((2+i\mu+i\nu-\veps)/2)\Gamma((2-i\mu+i\nu-\veps)/2)
\Gamma((2+i\mu-i\nu-\veps)/2)} \cr
&+ \left . {(e^{\pi(\mu+\nu)}+1) \Gamma((i\mu-i\nu+\veps)/2) \over
\Gamma((2+i\mu-i\nu-\veps)/2)\Gamma((2-i\mu-i\nu-\veps)/2)
\Gamma((2+i\mu+i\nu-\veps)/2)}\right ]\ .
\label{i.5}\end{align}
These expressions have well-defined limits in the sense of distributions in
$\mu$ and $\nu$.
In the numerator of each term inside the square brackets we make
the substitution
$\Gamma(z) = \Gamma(1+z)/z$. As $\veps \rightarrow 0$, we find
\beqa
&& \wt{K_{\mu,\nu}^{u,\veps}}(\kr) + \wt{K_{\mu,\nu}^{d,\veps}}(\kr)
\sim {2^{d-1}\pi^{d+1} k^{i(\mu-\nu)}\over
\sh (\pi \mu)\,\sh (\pi \nu)\,
\Gamma\left(\frac{d-1}{2} - i\,\mu \right)
\Gamma\left(\frac{d-1}{2} + i\,\nu
\right)}
\times \cr
&&\left [ {e^{\pi(\mu+\nu)}+1 \over
\Gamma\left(1+i{\mu+\nu\over 2}\right )
\Gamma\left(1-i{\mu+\nu\over 2}\right )}\left (
{1\over i(\mu-\nu)+\veps}+{1\over -i(\mu-\nu)+\veps}\right )
\right .\cr
&&\left . -{e^{\pi\mu}+e^{\pi\nu} \over
\Gamma\left(1+i{\mu-\nu\over 2}\right )
\Gamma\left(1-i{\mu-\nu\over 2}\right )}\left (
{1\over i(\mu+\nu)+\veps}+{1\over -i(\mu+\nu)+\veps}\right )
\right ]\ .
\label{i.6}\endqa
Using $(it+\veps)^{-1}+ (-it+\veps)^{-1} \sim 2\pi\,\delta(t)$,
and $\Gamma(1+iz)\Gamma(1-iz) = \pi z/\sh(\pi z)$,
this gives
\beqa
&& \wt{K_{\mu,\nu}^{u,\veps}}(\kr) + \wt{K_{\mu,\nu}^{d,\veps}}(\kr)
\sim {2^d\pi^{d+1} k^{i(\mu-\nu)}\over
\sh (\pi \mu)
\Gamma\left(\frac{d-1}{2} - i\,\mu \right)
\Gamma\left(\frac{d-1}{2} + i\,\nu
\right)}
\times \cr
&&\left [ {e^{2\pi\mu}+1 \over \mu}\,\delta(\mu-\nu)
+ {e^{\pi\mu}+e^{-\pi\mu} \over \mu}\,\delta(\mu+\nu) \right ]\ .
\label{i.7}\endqa
Recall that we are interested in the case when $\mu\not=0$ and
$\nu \not=0$ have the same sign. In this case $\delta(\mu+\nu)=0$, and
$|\mu|^{-1}\delta(\mu-\nu) = 2\delta(\mu^2-\nu^2)$. Thus, in this case,
\beq
\wt{K_{\mu,\nu}^{u,\veps}}(\kr) + \wt{K_{\mu,\nu}^{d,\veps}}(\kr) \sim
{2^{d+2}\pi^{d+1} e^{\pi\mu}|\coth(\mu)|\over
\Gamma\left(\frac{d-1}{2} - i\,\mu \right)
\Gamma\left(\frac{d-1}{2} + i\,\mu \right)}\,\delta(\mu^2-\nu^2)\ .
\label{i.8}\endq
Therefore, by (\ref{i.2}) and (\ref{i.2.1}),
\beqa
\lefteqn{
K_{\mu,\nu}(\xi,\xi',g=1) =
K_{\mu,\nu}^{u}(\xi,\ \xi') + K_{\mu,\nu}^{d}(\xi,\ \xi') =}\cr
&& {2^{d+2}\pi^{d+1} e^{\pi\mu}|\coth(\mu)|\over
\Gamma\left(\frac{d-1}{2} - i\,\mu \right)
\Gamma\left(\frac{d-1}{2} + i\,\mu \right)}\,
\delta(\mu^2-\nu^2)\,\delta(\eta-\eta')\ ,
\label{i.9}\endqa
Recall that this holds when $\mu$ and $\nu$ are both non-zero
and have the same sign. Still in the same case, using (\ref{f.16.2})
we have
\beqa
&& \int_{X_d} \w_\mu(x,\ y)\w_\nu(y,\ x')\,dy =\cr
&& c_{d,\mu}c_{d,\nu}\int_{\gamma\times \gamma}
\psi_{-{d-1\over 2}+i\mu}(x,\ \xi)\,K_{\mu,\nu}(\xi,\xi',1)\,
\psi_{-{d-1\over 2}-i\nu}(x',\ \xi')\,d\mu_\gamma(\xi)\,
d\mu_\gamma(\xi')\ .
\label{i.11}\endqa
We choose $\gamma=V_0$ as described at the beginning of Appendix
\ref{adpar} and of this Appendix. With the parametrization (\ref{ap.1}),
this is a $(d-1)$-Euclidean space and $d\mu_\gamma(\xi) = d^{d-1}\eta$.
Therefore, by (\ref{i.9}), the rhs of (\ref{i.11}) is given by
\beq
(c_{d,\mu})^2 \delta(\mu^2-\nu^2)
{2^{d+2}\pi^{d+1} e^{\pi\mu} |\coth(\mu)| \over
\Gamma\left(\frac{d-1}{2} - i\,\mu \right)
\Gamma\left(\frac{d-1}{2} + i\,\mu \right)}
\int_\gamma \psi_{-{d-1\over 2}+i\mu}(x,\ \xi)\,
\psi_{-{d-1\over 2}-i\mu}(x',\ \xi)\,d\mu_\gamma(\xi),
\label{i.12}\endq
and finally
\beq
\int_{X_d} \w_\mu(x,\ y)\w_\nu(y,\ x')\,dy =
2\pi |\coth(\mu)| \delta(\mu^2-\nu^2)\,\w_\mu(x,\ x')\ .
\label{i.13}\endq
Although we have assumed $\mu$ and $\nu$ to have the same sign in
the derivation, it follows from $\w_\mu = \w_{-\mu}$ and the form
of the formula above that it holds for all possible relative
signs, provided $\mu\not=0$ and $\nu\not=0$.

\bibliographystyle{unsrt}

\begin{thebibliography}{10}

\bibitem{Riess:1998cb}
Adam~G. Riess et~al.
\newblock {Observational Evidence from Supernovae for an Accelerating Universe
  and a Cosmological Constant}.
\newblock {\em Astron. J.}, 116:1009--1038, 1998.

\bibitem{Perlmutter:1998np}
S.~Perlmutter et~al.
\newblock {Measurements of Omega and Lambda from 42 High-Redshift Supernovae}.
\newblock {\em Astrophys. J.}, 517:565--586, 1999.

\bibitem{desitter}
W.~De~Sitter.
\newblock On the relativity of inertia: remarks concerning {Einstein}'s latest
  hypothesis.
\newblock {\em Proc. Kon. Ned. Acad. Wet.}, 19:1217--1225, 1917.

\bibitem{desitterbis}
W.~De~Sitter.
\newblock On the curvature of space.
\newblock {\em Proc. Kon. Ned. Acad. Wet.}, 20:229--243, 1917.

\bibitem{N}
O.~Nachtmann.
\newblock Dynamische {S}tabilit{\"a}t im de-{S}itter-raum.
\newblock {\em Osterr. Akad. Wiss., Math.-Naturw. Kl.}, Abt. II 176:363--379,
  1968.

\bibitem{Boyanovsky:1996ab}
D.~Boyanovsky, R.~Holman, and S.~Prem~Kumar.
\newblock {Inflaton decay in De Sitter spacetime}.
\newblock {\em Phys. Rev.}, D56:1958--1972, 1997.

\bibitem{Boyanovsky:2004gq}
Daniel Boyanovsky and Hector~J. de~Vega.
\newblock {Particle decay in inflationary cosmology}.
\newblock {\em Phys. Rev.}, D70:063508, 2004.

\bibitem{Boyanovsky:2004ph}
Daniel Boyanovsky, Hector~J. de~Vega, and Norma~G. Sanchez.
\newblock {Particle decay during inflation: Self-decay of inflaton quantum
  fluctuations during slow roll}.
\newblock {\em Phys. Rev.}, D71:023509, 2005.

\bibitem{Boyanovsky:2003ui}
D.~Boyanovsky and H.~J. de~Vega.
\newblock {Dynamical renormalization group approach to relaxation in quantum
  field theory}.
\newblock {\em Ann. Phys.}, 307:335--371, 2003.

\bibitem{Bros:2006gs}
J.~Bros, H.~Epstein, and U.~Moschella.
\newblock {Lifetime of a massive particle in a de Sitter universe}.
\newblock {\em JCAP}, 0802:003, 2008.

\bibitem{Gursey}
F.~G{\"u}rsey.
\newblock Introduction to the de {S}itter group.
\newblock In {\em Group theoretical concepts and methods in elementary particle
  physics}, pages 365--389. Gordon and Breach, New York, 1964.

\bibitem{Gibbons:1977mu}
G.~W. Gibbons and S.~W. Hawking.
\newblock {Cosmological Event Horizons, Thermodynamics, and Particle Creation}.
\newblock {\em Phys. Rev.}, D15:2738--2751, 1977.

\bibitem{Bros:1994dn}
J.~Bros, U.~Moschella, and J.~P. Gazeau.
\newblock {Quantum field theory in the de Sitter universe}.
\newblock {\em Phys. Rev. Lett.}, 73:1746--1749, 1994.

\bibitem{Bros:1995js}
J.~Bros and U.~Moschella.
\newblock {Two-point Functions and Quantum Fields in de Sitter Universe}.
\newblock {\em Rev. Math. Phys.}, 8:327--392, 1996.

\bibitem{Bros:1998ik}
J.~Bros, H.~Epstein, and U.~Moschella.
\newblock {Analyticity properties and thermal effects for general quantum field
  theory on de Sitter space-time}.
\newblock {\em Commun. Math. Phys.}, 196:535--570, 1998.

\bibitem{BLOT}
N.N. Bogolubov, A.A. Logunov, A.I. Oksak, and I.T. Todorov.
\newblock {\em {General principles of Quantum Field Theory}}.
\newblock {Springer-Verlag}, {Berlin Heidelberg New York}, 1990.

\bibitem{bv}
J.~Bros and G.~A. Viano.
\newblock {\em Forum Math.}, 8:621, 1996.

\bibitem{jost}
R.~Jost.
\newblock {\em {The general theory of quantized fields}}.
\newblock A.M.S., Providence, RI, 1965.

\bibitem{araki}
Huzihiro Araki.
\newblock {\em {Mathematical theory of quantum fields}}.
\newblock Oxford University Press, Oxford, 1999.

\bibitem{Bros:1996mw}
Jacques Bros and Detlev Buchholz.
\newblock {Axiomatic analyticity properties and representations of particles in
  thermal quantum field theory}.
\newblock {\em Ann. Poincare}, 64:495--522, 1996.

\bibitem{Veltman}
M.~Veltman.
\newblock {\em {Diagrammatica}}, volume~I.
\newblock Cambridge University Press, Cambridge, 1994.

\bibitem{kayw}
B.~S. Kay and R.~M. Wald.
\newblock {Theorems on the Uniqueness and Thermal Properties of Stationary,
  Nonsingular, Quasifree States on Space-Times with a Bifurcate Killing
  Horizon}.
\newblock {\em Phys. Rept.}, 207:49--136, 1991.

\bibitem{BEGMP}
J.~Bros, H.~Epstein, M.~Gaudin, U.~Moschella, and V.~Pasquier.
\newblock {Triangular invariants, three-point functions and particle stability
  on the de Sitter universe}.
\newblock {\em In preparation}.

\bibitem{MagnusOS}
W.~Magnus, F.~Oberhettinger, and R.~P. Soni.
\newblock {\em {Formulas and Theorems for the Special Functions of mathematical
  Physics}}.
\newblock Springer-Verlas, Berlin Heidelberg New York, 1966.

\bibitem{Marichev}
O.I. Marichev.
\newblock {\em Handbook of Integral Transforms of Higher Transcendental
  Functions}.
\newblock Ellis Horwood Limited, Chichester, 1982.

\bibitem{Slater}
L.J. Slater.
\newblock {\em Generalized Hypergeometric Functions}.
\newblock Cambridge University Press, Cambridge, 1966.

\bibitem{bateman1}
A.~Erd\'elyi.
\newblock {\em {The Bateman manuscript project. Higher Transcendental
  Functions}}, volume~I.
\newblock McGraw-Hill, New York, 1953.

\bibitem{skenderis}
K.~Skenderis and P.~K. Townsend.
\newblock {Pseudo-supersymmetry and the domain-wall / cosmology
  correspondence}.
\newblock {\em J.\ Phys.\ A}, 40:6733, 2007.

\bibitem{Abbott:1981ff}
L.~F. Abbott and S.~Deser.
\newblock {Stability of Gravity with a Cosmological Constant}.
\newblock {\em Nucl. Phys.}, B195:76, 1982.

\bibitem{Vi}
N.Ja. Vilenkin.
\newblock {\em Special Functions and the Theory of Group Representations}.
\newblock Nauka, Moscow, 1968.

\bibitem{bateman2}
A.~Erd\'elyi.
\newblock {\em {The Bateman manuscript project. Higher Transcendental
  Functions}}, volume~II.
\newblock McGraw-Hill, New York, 1953.

\end{thebibliography}

\end{document}

